\documentclass[a4paper,twocolumn,reprint,prx,amsfonts]{revtex4-1}
\usepackage[utf8]{inputenc}
\usepackage{color}
\usepackage{verbatim}
\usepackage{amsmath}
\usepackage{amssymb}
\usepackage{graphicx}
\usepackage[unicode=true,
 bookmarks=false,
 breaklinks=false,pdfborder={0 0 1},backref=false,colorlinks=true]
 {hyperref}
\hypersetup{pdftitle={`Free-Space' Photonic Quantum Link and Chiral Quantum Optics},
 pdfauthor={A. Grankin, P.\,O. Guimond D. V. Vasilyev, B. Vermersch, P. Zoller},
 linkcolor=blue,citecolor=blue,urlcolor=blue,linkcolor=blue,citecolor=blue,urlcolor=blue}

\makeatletter

\pdfpageheight\paperheight
\pdfpagewidth\paperwidth

\usepackage{braket}

\definecolor{mygreen}{rgb}{0,0.5,0}
\definecolor{myblue}{rgb}{0,0,0.75}
\definecolor{mymagenta}{cmyk}{0,1,0,0.12}

\makeatother

\begin{document}

\title{`Free-Space' Photonic Quantum Link and Chiral Quantum Optics}

\author{A.~Grankin}
\author{P.\,O.~Guimond}
\author{D.\,V.~Vasilyev}
\author{B.~Vermersch}
\author{P.~Zoller}
\affiliation{Institute for Theoretical Physics, University of Innsbruck} 
\affiliation{Institute
for Quantum Optics and Quantum Information, Austrian Academy of Sciences,
Innsbruck, Austria}

\begin{abstract}
We present the design of a chiral photonic quantum link, where distant atoms interact by exchanging photons propagating in a single direction in free-space. This is achieved by coupling each atom in a laser-assisted process to an atomic array acting as a quantum phased-array antenna. This provides a basic building block for quantum networks in free space, i.e. without requiring cavities or nanostructures, which we illustrate with high-fidelity quantum state transfer protocols. Our setup can be implemented with neutral atoms using Rydberg-dressed interactions.
\end{abstract}
\maketitle

\section{Introduction}

Modular architectures for quantum information processing envision a scale-up of quantum devices from elementary building blocks interconnected by coherent quantum links \cite{Kimble2008,Duan2010,Nickerson2014,Monroe2014}. A modular quantum processor is structured as an `on-chip' local-area quantum network, with small quantum computers as nodes of the network, and where quantum states are transferred between nodes via quantum channels \cite{Northup2014,Reiserer2015}. Remarkable progress has been made during the last years in demonstrating some of the basic elements of such modular architectures. In atomic physics quantum computers and quantum simulators involving several tens of individually controlled qubits have been built along with quantum logic entangling operations \cite{Labuhn2016,Bernien2017,Zhang2017,Friis2018}. Deterministic and probabilistic protocols for entanglement generation \cite{Haas2014}, as well as quantum state transfer between distant atomic qubits \cite{Ritter2012} have been demonstrated experimentally using cavity-QED interfaces, with optical photons as carriers of quantum information.  A basic requirement and remaining challenge, however, is to develop high-speed  photonic quantum links allowing for high-fidelity quantum communication and entanglement distribution in on-chip quantum networks. 

\begin{figure*}[t]
\includegraphics[width=1\linewidth]{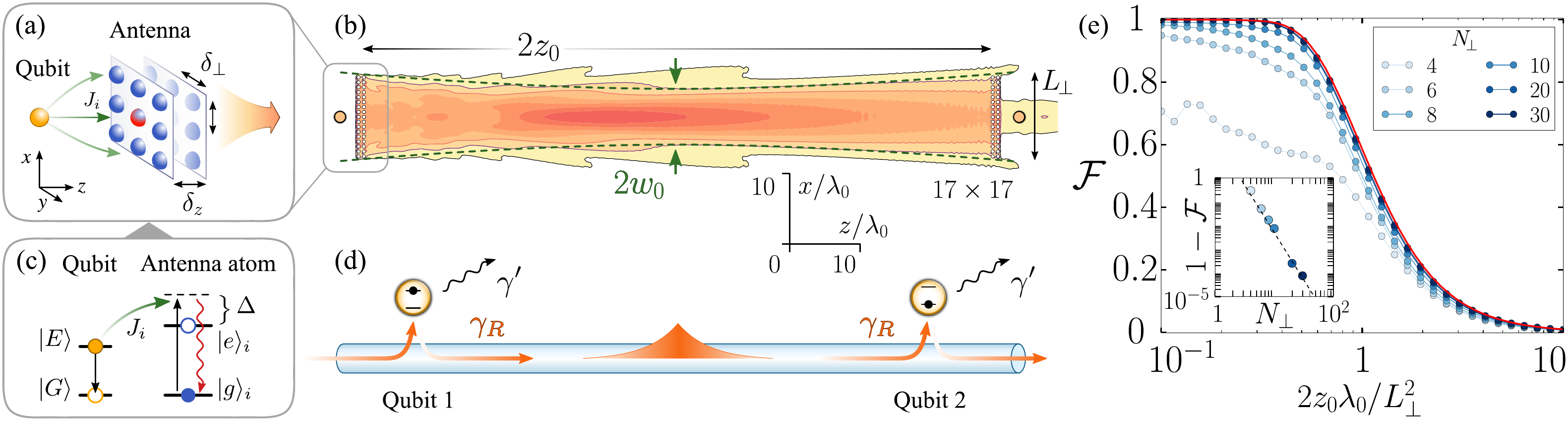} \caption{\textit{Free-space chiral quantum optics.} (a)~A master atom (qubit) is coupled 
to a bilayer atomic array of $N_{\rm a}=N_\perp\times N_\perp \times 2$ atoms as `quantum antenna' achieving
unidirectional photon emission. (b)~Spatial distribution of the emitted
field $|\vec{\varphi}(\vec{r}\,)|$ (see text) for two master atoms as qubits interacting via a paraxial mode, with $2z_0=90\lambda_0$, $N_\perp=17$, $\delta_\perp=0.7\lambda_0$. 
 (c)~Basic model
of the antenna: level schemes of the qubit off-resonantly coupled
to a two-level antenna atom. (d) Chiral coupling in waveguide-QED, with two atomic qubits coupling to right-propagating modes with rate $\gamma_{R}$, and to non-guided modes with rate ${\gamma'}$. (e) Fidelity for Quantum State Transfer between atomic qubits, with interatomic spacing $\delta_\perp=0.8\lambda_0$, and various $N_\perp$. The red curve corresponds to $N_\perp\to\infty$ (see text). Inset: corresponding infidelity for $z_0\to0$, with $50/N_\perp^4$ in dashed black for reference. }
\label{fig:chiralQO} 
\end{figure*}

The paradigmatic and conventional setup of a photonic quantum link is built around strong coupling of atoms to photonic nanostructures or nanofibers as 1D waveguides \cite{Tiecke2014,Hood2016,Corzo2016,Solano2017}, or to cavities \cite{Ritter2012,McConnell2015}. In Fig.~\ref{fig:chiralQO}(d), we sketch an example of photonic quantum link between two atomic qubits, based on an interface between a two-level atom and an optical fiber. In such setups protocols can be applied to deterministically transfer a quantum state from the first to the second atomic `stationary' qubit via a photonic `flying' qubit propagating as wavepacket in a 1D optical waveguide. Achieving a high-fidelity transfer requires the following two key ingredients. First, we need to achieve routing of the photon wavepacket emitted by the first atom. This necessitates a \textit{chiral} atom-fiber interface, i.e. with unidirectional photon emission (and absorption) \cite{Lodahl2017}.  The second ingredient is the 1D character of the fiber modes guiding the wavepacket, which is essential in achieving efficient reabsorption of the photon and thus restoration of the qubit in the second atom.  Recent experiments have demonstrated such {chiral} quantum interfaces with atoms trapped close to optical nanofibers~\cite{Mitsch2014,LeFeber2015}. Significant challenges remain, however, in resolving the conflicting requirements of trapping atoms close to dielectric surfaces, while achieving the strong-coupling regime where the interaction between atoms and confined modes dominates losses such as spontaneous emission to free-radiating modes. These challenges have so far limited demonstrations of photon mediated remote entanglement of matter qubits
to rates of at most 30 $s^{-1}$ with neutral atoms \cite{Ritter2012}, 
trapped ions \cite{Hucul2014}, or NV centers \cite{Kalb2017}. To address these challenges, we propose below a chiral photonic quantum link, where effective 1D (paraxial) \textit{free-space} modes of the electromagnetic field provide a photonic quantum channel connecting atomic qubits --- therefore eliminating the requirement for 1D nanofibers or photonic nanostructures. Combining the demonstrated capability for creating local entanglement on a few $\mu$s timescale via
Rydberg interactions \cite{Maller2015,Jau2015} with a high efficiency chiral channel 
we project realization of remote entanglement at rates above $10^4~\rm s^{-1}$, which will speed 
development of modular quantum processing networks.  

The setup describing the `free-space' photonic quantum link is outlined in Figs.~\ref{fig:chiralQO}(a-c). The key element is the coupling of the atom representing the qubit, in an engineered laser-assisted process, to a regular array of atoms with sub-wavelength separation $\delta_\perp<\lambda_0$ (with $\lambda_0$ the wavelength of the light), which acts as a \textit{phased-array} antenna for photon emission and absorption. We make this interface \textit{chiral} by employing a bilayer atomic array, where the desired unidirectionality is guaranteed by interference. We can view the composite object consisting of the qubit atom coupled to the atomic array as an  artificial two-level atom, where the `excited state' decays to the `ground state' while coherently emitting an optical photon into a given well-defined localized and directed (1D) mode of the electromagnetic field. This chiral photonic quantum interface for `free-space' atomic qubits then becomes the building block for a `free-space' photonic quantum link. We illustrate this in Fig.~\ref{fig:chiralQO}(c) for the example of an array of $N_\perp \times N_\perp$ atoms (here $17 \times 17$)  with $N_z=2$ layers acting as antenna. This example demonstrates the generation of a free-space Gaussian mode as photonic quantum link connecting two atomic arrays. For a given transverse array size $L_\perp \sim  \lambda_0 N_\perp$, this link can cover a distance $L\sim L_\perp ^2 /\lambda_0 \sim N_\perp L_\perp$ between sending and  receiving node.
The achievable communication range can be further extended with lenses inserted between the sending and receiving antenna.
Remarkably, running the standard quantum state transfer protocol on this setup gives fidelities close to unity for such distances, as shown in Fig.~\ref{fig:chiralQO}(e). For atomic arrays of much smaller size, the light emitted from the antenna remains unidirectional, albeit divergent as illustrated in Fig.~\ref{fig:setup}(a).% [see Fig.~\ref{fig:setup}(b)].

\section{Chiral quantum optics}

\label{sec:chiralQO}

\label{sec:chiralQO}

We wish to implement a `free-space' chiral light-matter
interface (discussed in Sec.~\ref{sec:Model}), and a `free-space' photonic quantum link
(discussed in Sec.~\ref{sec:chiralnetworks}) in a 3D environment, analogous to the 1D models of
chiral quantum optics. Thus, for reference below, we find it worthwhile
to first summarize the basic dynamical equations of 1D chiral quantum optics
and cascaded quantum systems. 

A minimal model for a chiral interface
coupling two-level atoms to a waveguide is shown in Fig.~\ref{fig:chiralQO}(d).
Here, two quantum emitters %
\mbox{%
$(a=1,2)$%
} as two-level atoms with ground states $\ket{G}_{a}$ and excited
states $\ket{E}_{a}$, respectively, are coupled to an open 1D waveguide
as bosonic bath.  The dynamics of this system is governed by the Hamiltonian $H_{1D}=H_{0A}+H_{0F}+H_{AF}$.
Here the free Hamiltonian for the waveguide can be written as %
\mbox{%
$H_{0F}=\int dk\omega_{k}\left({b_{k}^{R}}^{\dagger}b_{k}^{R}+{b_{k}^{L}}^{\dagger}b_{k}^{L}\right)$%
}, where $\omega_{k}$ is the waveguide dispersion relation, which
we assume linear ($\omega_{k}\approx ck$ with $k$ the momentum and
$c$ the speed of light in the waveguide), and ${b_{k}^{R(L)}}$ is
the annihilation operator for photons propagating in the right (left)
direction in the waveguide, with momentum $k$, which satisfy bosonic
commutation relations %
\mbox{%
$\left[b_{k}^{\alpha},{b_{k'}^{\beta}}^{\dagger}\right]=\delta(k-k')\delta_{\alpha,\beta}$%
}. On the other hand, the free atomic Hamiltonian is %
\mbox{%
$H_{0A}=\omega_{0}\sum_{a=1}^{2}\sigma_{a}^{+}\sigma_{a}^{-}$%
}, with $\omega_{0}$ the atomic transition frequency and $\sigma_{a}^{-}\equiv\ket{G}_{a}\!\bra{E}$,
possibly along with an additional term accounting for external driving
fields. Finally, the interaction Hamiltonian between atoms and photons
reads 
\begin{equation}
\begin{aligned}H_{AF}=i & \sum_{a=1}^{2}\sqrt{\frac{\gamma_{R,a}}{2\pi}}\int dk\left(e^{-ikz_{a}}{b_{k}^{R}}^{\dagger}\sigma_{a}^{-}-\text{h.c.}\right)\\
+ & i\sum_{a=1}^{2}\sqrt{\frac{\gamma_{L,a}}{2\pi}}\int dk\left(e^{ikz_{a}}{b_{k}^{L}}^{\dagger}\sigma_{a}^{-}-\text{h.c.}\right),
\end{aligned}
\end{equation}
where $z_{a}$ is the atomic position along the waveguide, with $d\equiv z_{2}-z_{1}>0$,
and $\gamma_{R(L),a}$ is the \emph{spontaneous decay rate} of atom
$a$ for the emission of photons propagating to the right (left).
Broken left-right symmetry manifests itself in the couplings $\gamma_{R,a}\ne\gamma_{L,a}$,
and we are particularly interested in unidirectional coupling $\gamma_{R,a}\gg\gamma_{L,a}\rightarrow0$.

The master equation obtained by integrating out the radiation field
in a Born-Markov approximation for two atoms and $\gamma_{R,a}\ne\gamma_{L,a}$
is
\begin{equation}
\frac{d}{dt}{\rho}=-i\left[H_{\text{eff}}{\rho}-{\rho}H_{\text{eff}}^{\dagger}\right]+\mathcal{J}{\rho},\label{eq:mastereq1D}
\end{equation}
with non-hermitian Hamiltonian 
\begin{equation}
H_{\text{eff}}=-i\sum_{a=1}^{2}\frac{\gamma_{a}}{2}s_{a}^{+}s_{a}^{-}-ie^{i\omega_{0}d/c}\Big(\gamma_{L}s_{1}^{+}s_{2}^{-}+\gamma_{R}s_{2}^{+}s_{1}^{-}\Big).\label{eq:Hchir1D}
\end{equation}
written here in a rotating frame. The first term in Eq.~(\ref{eq:Hchir1D}) describes the individual decay of atomic excitations, with the total
decay rate of atom $a$ defined as %
\mbox{%
$\gamma_{a}=\gamma_{R,a}+\gamma_{L,a}+\gamma'_{a}$%
}. Here we added an additional decay channel with rate $\gamma'_{a}$
accounting for losses due to coupling of the atoms to non-guided modes.
The second term of Eq.~(\ref{eq:Hchir1D}) on the other hand describes
non-reciprocal atomic effective interactions. The rate \mbox{$\gamma_{L}\equiv\sqrt{\gamma_{L,1}\gamma_{L,2}}$}
denotes the rate of interaction mediated by photons propagating to
the left from atom $2$ to $1$, while $\gamma_{R}\equiv\sqrt{\gamma_{R,1}\gamma_{R,2}}$
corresponds to photons propagating to the right from atom $1$ to
$2$. Finally, the last term in Eq.~(\ref{eq:mastereq1D})
expresses as
\begin{eqnarray}
\mathcal{J}{\rho}= &  & \sum_{a=1}^{2}\gamma_{a}s_{a}^{-}{\rho}s_{a}^{+}+e^{i\omega_{0}d/c}\left(\gamma_{R}s_{1}^{-}{\rho}s_{2}^{+}+\gamma_{L}s_{2}^{-}{\rho}s_{1}^{+}\right)\nonumber \\
 &  & +e^{-i\omega_{0}d/c}\left(\gamma_{L}s_{1}^{-}{\rho}s_{2}^{+}+\gamma_{R}s_{2}^{-}{\rho}s_{1}^{+}\right).
\end{eqnarray}

In the unidirectional case ($\gamma_{L}=0$), the above equation reduces
to the cascaded master equation as derived in Ref.~\cite{Gardiner1993}. We note that in
this case atom 1 can only talk to atom 2 downstream, while there is
no backaction of atom 2 to atom 1. This cascaded master equation has
been the starting point to discuss quantum state transfer of a qubit
as superposition state, from the first to the second atom, realizing %
 \mbox{%
$\left(\alpha\ket{G}_{1}+\beta\ket{E}_{1}\right)\otimes\ket{G}_{2}\to\ket{G}_{1}\otimes\left(\alpha\ket{G}_{2}+\beta\ket{E}_{2}\right)$%
} \cite{Cirac1997}.

We show below that the `free-space' chiral photonic quantum link of
Sec.~\ref{sec:chiralnetworks} can be described by a chiral master equation of the form of Eq.~\eqref{eq:mastereq1D} and we derive
explicit expressions for $\gamma_{R,a}\gg\gamma_{L,a}$ in terms of coupling coefficients to free-space radiation modes. The
setup of Figs.~\ref{fig:chiralQO}(a-c) thus provides a faithful implementation of chiral
quantum optics in a free space environment. 

\section{`Free-space' Chiral Atom-Light Interface\label{sec:Model}}
The basic setup of an atom coupled to a quantum antenna as directional
quantum emitter is illustrated in Figs.~\ref{fig:chiralQO}(a-c).
We consider a two-level atom represented by a pair of long lived atomic
states $\ket{G},\ket{E}$ (e.g.~hyperfine states in an atomic ground
state manifold), dubbed 'master atom' or qubit, which we assume trapped
in free space. We wish to design an effective `decay' from the excited
state to the ground state $\ket{E}\rightarrow\ket{G}$ as a laser-assisted
spontaneous emission process, analogous to an optical pumping process,
with the property that the optical photon is emitted into a specified
target mode of the electromagnetic field, written as an outgoing wave
packet \mbox{$\left\vert \psi^{\text{targ}}\left(t\right)\right\rangle \equiv\sum_{\lambda}\int d^{3}k\psi_{\vec{k},\lambda}^{\text{targ}}(t)b_{\vec{k},\lambda}^{\dagger}\ket{\textrm{vac}}$},
with $\ket{\textrm{vac}}$ the vacuum state. Here $b_{\vec{k},\lambda}^{\dagger}$
creates a photon with momentum $\vec{k}$ and polarization $\lambda$,
with \mbox{$[b_{\vec{k},\lambda},b_{\vec{k}'\lambda'}^{\dagger}]=\delta_{\lambda,\lambda'}\delta(k-k')$},
and $\psi_{\vec{k},\lambda}^{\text{targ}}$ specifies the target mode
in momentum space (e.g.~a Gaussian mode).

We design this `decay' of the master atom with the optical photon
emitted into the target mode as a two-step process via a nearby atomic array with subwavelength spacing (as investigated in recent theoretical studies \cite{Bettles2016,Facchinetti2016,Shahmoon2017,Asenjo-Garcia2017,Manzoni2017,Perczel2017}). This ensemble consists
of two-level atoms $\left\{ \ket{g}_{i},\ket{e}_{i}\right\} $ located
at positions $\vec{r}_{i}$ ($i=1,...,N_{a}$) trapped in free-space (e.g. with optical traps \cite{Bloch2012,Lester2015,Xia2015,Endres2016,Barredo2016}). In a first step the
atomic excitation of the master atom is swapped in a laser assisted
process to a \emph{delocalized} electronic excitation of the ensemble,
\begin{equation}
s^{+}\ket{\Omega}\rightarrow\sum_{i=1}^{N_{a}}e^{i\phi_{i}}\sigma_{i}^{+}\ket{\Omega}/\sqrt{N_{a}}.\label{eq:transfer}
\end{equation}
Here $\ket{\Omega}\equiv\ket{G}\{\ket{g}_{i}\}\ket{\textrm{vac}}$,
and we have defined %
\mbox{%
$s^{+}\equiv\ket{E}\bra{G}$%
} and %
\mbox{%
$\sigma_{i}^{+}\equiv\ket{e}_{i}\bra{g}$%
}. This delocalized electronic excitation in the atomic ensemble then
decays back to the ground state $\ket{e}_{i}\rightarrow\ket{g}_{i}$
by emission of an optical photon. The key idea is to design phases
$\phi_{i}$ in the laser-assisted first step, so that the atomic ensemble
acts as a holographic or \textit{phased-array antenna} for directed spontaneous emission into the target mode (in analogy to classical phased-array antennas \cite{He2009,Sun2013,Stokes2015}). That is,
directionality of emission comes from interference between the emitting
atomic dipoles \cite{Clemens2003,Saffman2005,Scully2006,Miroshnychenko2013,Slepyan2013,Wiegner2015,Bhatti2016,Paris-Mandoki2017}. We first assume the system to operate in the Lamb-Dicke regime, such that photon recoil and motional effects, e.g. due to temperature, can be neglected \cite{gardiner2015}.
There are
various ways of implementing the process (\ref{eq:transfer}) in quantum
optics with atoms; as an example we discuss below a transfer with
long-range laser-assisted Rydberg interactions \cite{Saffman2010,Browaeys2016},
where spatially dependent phases $\phi_{i}$ can be written via laser
light, akin to synthetic gauge fields for cold atoms \cite{Goldman2016}.
We emphasize that the overall process preserves quantum coherence
and entanglement, such that, for instance, for an initial qubit superposition
state $c_{g}\left\vert G\right\rangle +c_{e}\left\vert E\right\rangle $
(with $c_{g}$ and $c_{e}$ complex numbers) the outgoing photonic
state will read $c_{g}\left\vert \text{vac}\right\rangle +c_{e}\left\vert \psi^{\text{targ}}\left(t\right)\right\rangle $. 

\begin{figure}[t]
\includegraphics[width=1\linewidth]{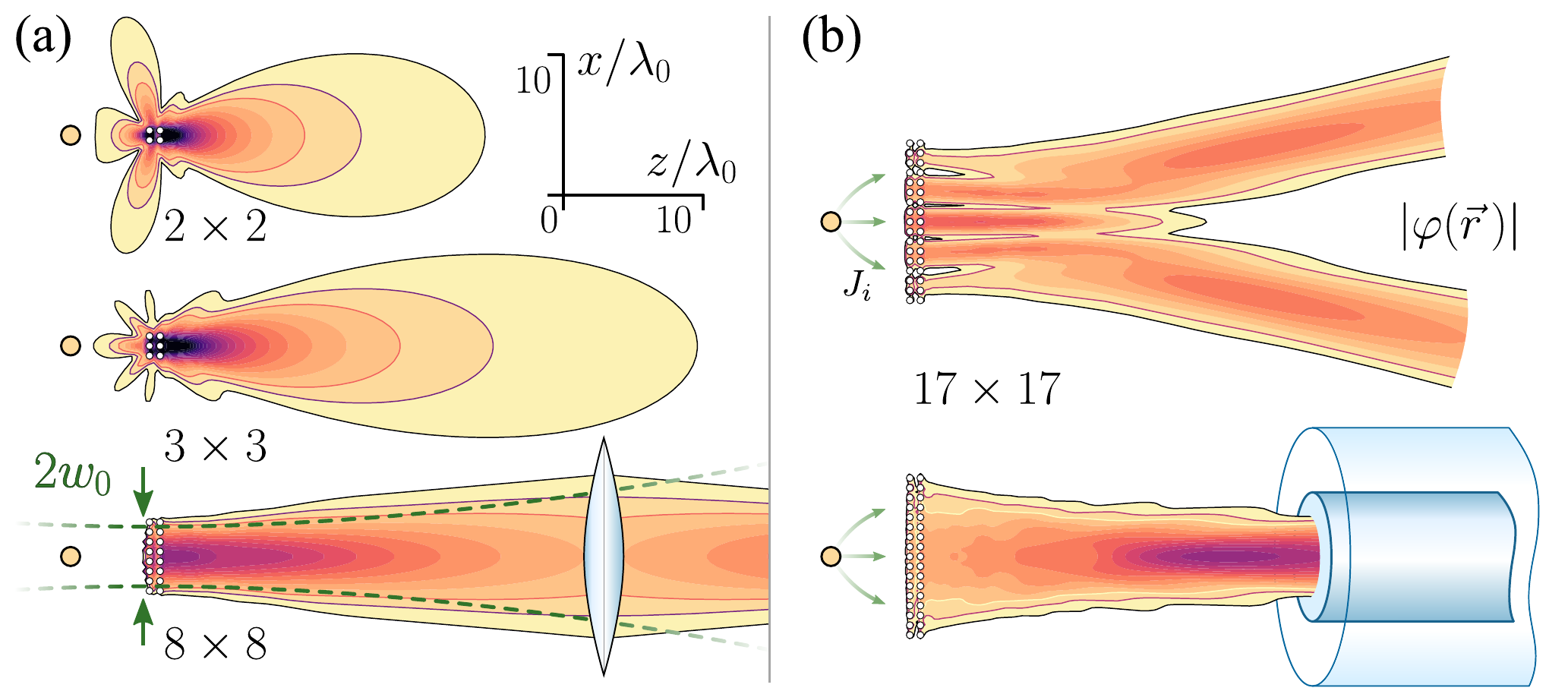} \caption{\textit{Few-atom quantum antenna.} (a)~Spatial distribution of the emitted
field $|\vec{\varphi}(\vec{r}\,)|$ (see text) for antenna configurations
as bilayer $\protect\smash{2 \times 2}$, $\protect\smash{3 \times 3}$
and $\protect\smash{8 \times 8}$ regular arrays. The field can be interfaced with optical lenses. (b)~Emission to a superposition of two
focussing modes, and mode matching to an optical fiber,
with bilayer $\protect\smash{17 \times 17}$ atomic arrays as antennas.}
\label{fig:setup} 
\end{figure}

Below we will be interested in various geometries of the few-atom
antenna, with the goal of optimizing the directionality of emission.
We will consider bilayer and multilayer regular arrays of $N_{{\rm a}}$
atoms, where %
\mbox{%
$N_{{\rm a}}=N_{\perp} \times N_{\perp} \times N_{z}$%
} with $N_{z}$ and $N_{\perp}$ being the number of atoms in longitudinal
and transversal directions, respectively. The corresponding interatomic
spacings are denoted as $\delta_{z(\perp)}$, while the overall spatial
extent of the antenna is $L_{z(\perp)}=\delta_{z(\perp)}N_{z(\perp)}$.
For comparison, we also assess the case of atoms with random positions
characterized by their density $n_{{\rm a}}$. As an illustration
of results derived below, we show in Fig.~\ref{fig:setup}(a) spatial
photon emission patterns for ${2\times2\times2}$, ${3\times3\times2}$
and ${8\times8\times2}$ bilayer regular arrays, assuming
subwavelength spacings $\delta_{z}=0.75\lambda_{0}$ and $\delta_{\perp}=0.7\lambda_{0}$, which is necessary in order to avoid Bragg resonances.
It is remarkable that rather directed spontaneous emission can be
obtained with very small atom numbers. We will quantify this below
as a Purcell factor $\beta$ for emission into a paraxial mode of
interest, and show that $\beta$ close to 1 can be achieved.

For transverse sizes $L_{\perp}\gg\lambda_{0}$, the antenna can emit
photons in several spatial modes, as illustrated in the upper panel
of Fig.~\ref{fig:setup}(b). Photons can also be emitted in directional
modes focussing at a distance $z_{0}$ outside the antenna, which
could be used to match the mode of an optical fiber, as represented
in the lower panel of Fig.~\ref{fig:setup}(b). The focussing distance
achievable in this way is limited by diffraction as $z_{0}\lesssim L_{\perp}^{2}/\lambda_{0}$.

\subsection{Model of Quantum Optical Antenna}

A quantum optical description for the setup in Fig.~\ref{fig:chiralQO}(a)
starts from a Hamiltonian \mbox{$H_{3D}=H_{0A}+H_{0F}+H_{AF}$}, which we write
as sum of an atomic Hamiltonian, the free radiation field, and the
atom - radiation field coupling in the dipole approximation. We find
it convenient to work in an interaction picture with respect to $H_{0F}$,
and transform to a rotating frame eliminating optical frequencies.
Thus we write for the atomic Hamiltonian (with $\hbar=1$) 
\begin{equation}
H_{0A}=-\Delta\sum_{i}\sigma_{i}^{+}\sigma_{i}^{-}+s^{-}\sum_{i}J_{i}\sigma_{i}^{+}+\text{h.c}.\label{eq:H0A}
\end{equation}
with $\Delta$ the detuning in the laser-assisted transfer from the
master atom to ensemble due to long-range couplings $J_{i}\!=\!|J_{i}|e^{i\phi_{i}}$.
For the atom-radiation field coupling we have 
\begin{equation}
H_{AF}(t)=-d\sum_{i}\sigma_{i}^{+}\vec{p}^{\,*}\vec{{\cal E}}^{\left(+\right)}\left(\vec{r}_{i},t\right)+\text{h.c.}\label{eq:HAF}
\end{equation}
with $\vec{d}= d \vec{p}$ the atomic dipole matrix element with amplitude $d$ and unit vector $\vec p$, which for concreteness we will assume circularly polarized along $z$, and
\begin{equation}\label{eq:expE}
\vec{{\cal E}}^{\left(+\right)}\left(\vec{r},t\right)=i\sum_{\lambda}\int d^{3}{k}\sqrt{\frac{\omega_{k}}{2(2\pi)^{3}\varepsilon_{0}}}b_{{\vec{k}},\lambda}e^{i{\vec{k}}{\vec{r}}}e^{-i(\omega_{\vec{k}}-\omega_{0})t}{\vec{e}}_{\lambda,{\vec{k}}}
\end{equation}
the positive frequency part of the electric field operator (in the
rotating frame), with $\omega_{0}\equiv ck_{0}\equiv2\pi c/\lambda_{0}$
the optical frequency, ${\vec{e}}_{\lambda,{\vec{k}}}$ the polarization
vector.

The effective decay of the master atom via the ensemble is described
in a Wigner-Weisskopf ansatz as 
\begin{equation}
\begin{aligned}\ket{\Psi({t})}=\Big( & \!s\left(t\right)s^{+}\!+\!\sum_{i}{\cal P}_{i}\left(t\right)\sigma_{i}^{+}\\
 & +\sum_{\lambda}\!\int d^{3}k\psi_{\vec{k},\lambda}\left(t\right)b_{k,\lambda}^{\dagger}\!\Big)\ket{\Omega},
\end{aligned}
\label{eq:WW}
\end{equation}
with initial condition $s(0)=1$, ${\cal P}_{i}\left(0\right)=\psi_{\vec{k},\lambda}\left(0\right)=0$,
and our aim is to obtain a photon in the specified target mode $\psi_{\vec{k},\lambda}\left(t\right)\rightarrow\psi_{\vec{k},\lambda}^{\text{targ}}(t)$
for times $t\rightarrow\infty$. In the Born-Markov approximation
we can eliminate the radiation field and find 
\begin{align}
\frac{d}{dt}s & =-i\sum_{i}J_{i}^{*}{\cal P}_{i},\quad\frac{d}{dt}{\cal P}_{i}=-iJ_{i}s-i\sum_{j}\left(H_{nh}\right)_{i,j}{\cal P}_{j}.\label{eq:S_P}
\end{align}
This describes the transfer of the excitation from the master atom
to the ensemble atoms according to the couplings $J_{i}$. The last
equation contains the non-hermitian effective atomic Hamiltonian \mbox{$H_{nh}\equiv-\Delta\mathbb{I}-i(\gamma_{e}/2)\left(\mathbb{I}+\mathbb{G}\right)$}
with detuning $\Delta$ and atomic decay rate $\gamma_{e}\equiv k_0^3 d^2/(3\pi\varepsilon_0)$, and the
hopping of the atomic excitation within the ensemble due to dipole-dipole
interaction induced by photon exchanges. Here we have defined 
\mbox{%
$\mathbb{G}_{i,j}\equiv\vec{p}^{\,*}\hat{G}\left(\vec{r}_{j}-\vec{r}_{i}\right)\vec{p}$%
}, where in our notations the dyadic Green's tensor is a solution of 
\mbox{$\vec{\nabla}\times\vec{\nabla}\times\hat{G}(\vec{r})-k_{0}^{2}\hat{G}(\vec{r})=-(6\pi i/k_0)\delta(\vec{r})\,\mathbb{I}$},
and expresses as \cite{Lehmberg1970,James1993} 
\begin{equation}
\begin{aligned}{\hat{G}}({\vec{r}})=\frac{3e^{ik_{0}r}}{2i(k_{0}r)^{3}}\Big[ & \left((k_{0}r)^{2}+ik_{0}r-1\right)\mathbb{I}\\
 & +\left(-(k_{0}r)^{2}-3ik_{0}r+3\right)\frac{\vec{r}\otimes\vec{r}}{r^{2}}\Big].
\end{aligned}
\end{equation}

To obtain the spatial profile of the emitted light, we define the
(normalized) single photon distribution as %
\mbox{%
$\vec{\psi}\left(\vec{r},t\right)\equiv-i\sqrt{{2\varepsilon_{0}}/{\omega_{0}}}\bra{\Omega}\vec{{\cal E}}^{\left(+\right)}\left(\vec{r}\right)\ket{\Psi\left(t\right)}$%
}. By integrating Maxwell equation we find 
\[
\vec{\psi}\left(\vec{{r}},t\right)=\sqrt{\frac{\gamma_{e}}{6\pi c}}k_{0}\sum_{i}\hat{G}\left(\vec{r}-\vec{r}_{i}\right)\vec{p}\,{\cal P}_{i}\left(t-\frac{|\vec{r}-\vec{r}_{i}|}{c}\right),
\]
with the emitted field as radiation of interfering atomic dipoles.

For simplicity we discuss below the limit of perturbative $J_{i}$,
where we eliminate the atomic ensemble coupled to the radiation field
as an effective quantum reservoir. We obtain for the effective decay
of the master atom 
\begin{align}
\frac{d}{dt}s & =\Big(i\sum_{i,j}{J_{i}}^{*}\left(H_{nh}^{-1}\right)_{i,j}{J_{j}}\Big)s\equiv\Big(-i\epsilon-\frac{1}{2}\gamma_{\text{tot}}\Big)s,\label{eq:full_gamma}
\end{align}
with $\gamma_{\text{tot}}$ the \emph{total emission rate} (into $4\pi$
solid angle). The spatio-temporal profile of the emitted photon wavepacket
can thus be written as $\vec{\psi}\left(\vec{r},t\right)=\vec{\varphi}(\vec{r}\,)s\left(\tau\right)$
with geometric factor 
\begin{equation}
\vec{\varphi}(\vec{r}\,)=-\sqrt{\frac{\gamma_{e}}{6\pi c}}k_{0}\sum_{i,j}\hat{G}\left(\vec{r}-\vec{r}_{i}\right)\vec{p}\left(H_{nh}^{-1}\right)_{i,j}J_{j}.\label{eq:phi}
\end{equation}
Here $s(\tau)=e^{-\gamma_{\text{tot}}\tau/2}$ represents the exponentially
decaying atomic state with retarded time $\tau=t\!-\!|\vec{r}\,|/c\ge0$.
We note that the above discussion generalizes to time-dependent couplings
$J_{j}\rightarrow J_{j}(t)\equiv J_{j}f(t)$, allowing a temporal
shaping of the outgoing wavepacket.

In our antenna design, we wish to optimize the directionality of emission
with an appropriate choice of phases $J_{j}=|J_{j}|e^{i\phi_{j}}$,
for a given antenna geometry and atomic parameters. In Figs.~\ref{fig:chiralQO} and \ref{fig:setup}
we have already presented corresponding results from numerical evaluation
of $\vec{\varphi}(\vec{r}\,)$ for various few-atom configurations.
We present both analytical and numerical studies of this optimization
problem in the following two sections.

\begin{figure*}[t]
\includegraphics[width=1\linewidth]{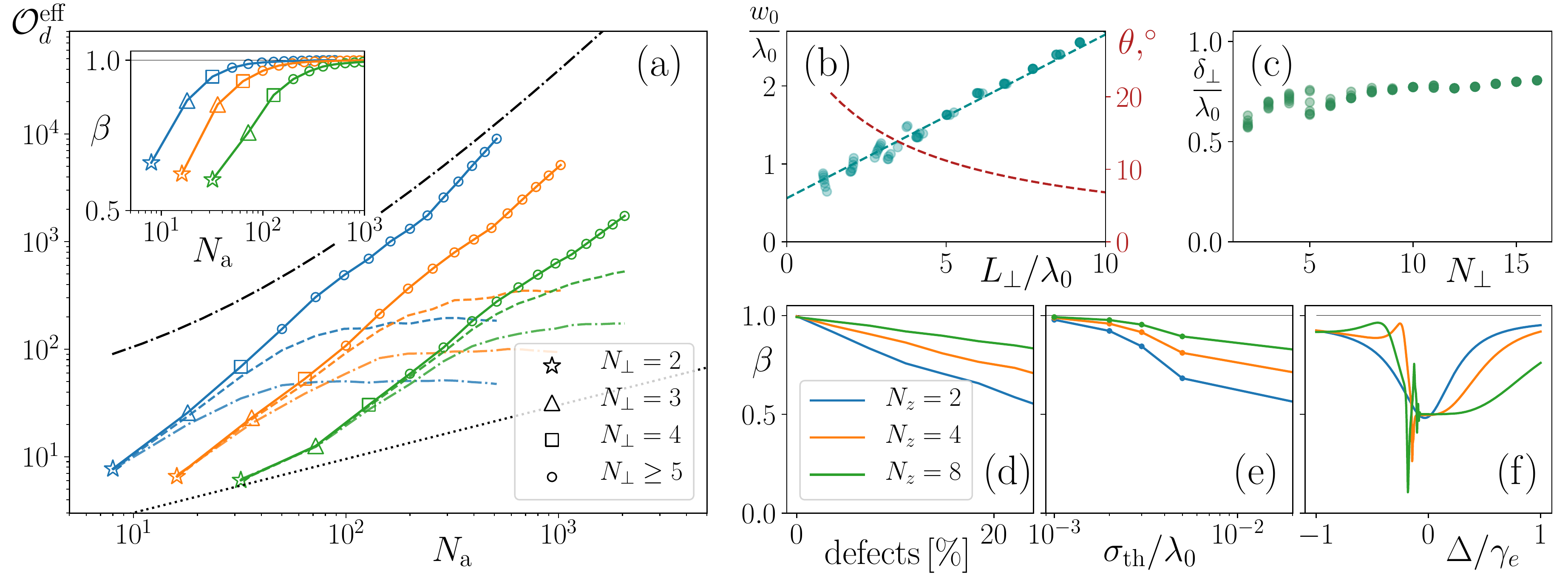}\caption{\emph{Performance of a `few-atom' antenna.} Effective optical depth
(a) and corresponding Purcell factor $\beta$ {[}inset~(a){]} for
an optimal Gaussian mode $n_{0}$, for an antenna consisting of $N_{{\rm a}}$
atoms. Regular arrays with $N_{z}=2,4,8$ are shown in blue, orange,
and green solid lines, respectively (transversal sizes of $N_{\perp}=2,3,4$
are highlighted by different markers as shown in the legend). A lattice
imperfection modeled as classical randomization of atomic positions
with normal distribution with dispersion  $\sigma_\text{th}=0.01\lambda_{0}$ ($\sigma_\text{th}=0.02\lambda_{0}$)
is shown in dashed (resp. dotted dashed) lines with the corresponding
color code. Results for randomly distributed atoms with the atomic
density of the regular arrays $n_{a}=2/\lambda_{0}^{3}$ are shown
with black dotted line. (b) Statistics of optimal Gaussian mode waists
$w_{0}$ (blue dots) versus $L_{\perp}$ for all configurations of
$N_{\perp}$ and $N_{z}$ up to $\protect\smash{16 \times 16 \times 8}$.
Fitted linear dependence (blue dashed line) of $w_{0}$ and the corresponding
opening angle $\theta$ of the Gaussian beam (in red). (c) Statistics
of optimal interatomic distances $\delta_{\perp}$ versus $N_{\perp}$.
(d-f) Imperfections for arrays with $N_{z}=2,4,8$ and $N_{\perp}=10$.
Purcell factor as a function of (d) percentage of defects, (e) Gaussian
dispersion $\sigma_\text{th}$ of randomized atomic positions, and (f) detuning
from resonance for two-level atoms. }

\label{fig2} 
\end{figure*}

\subsection{Quantum Antenna in Paraxial Approximation\label{subsec:Parxial}}

An analytical insight for optimizing emission to a given spatial mode
of the radiation field can be obtained in the paraxial approximation
for $\vec{\psi}(\vec{r},t)$. This approximation is valid for strongly
directional emission, and for atomic antenna configurations with $L_{\perp}\gg\lambda_{0}$
{[}as illustrated in Figs.~\ref{fig:setup}(b,d){]}. In the paraxial
description the target mode is specified as a desired paraxial mode
of interest, e.g. as a Laguerre-Gauss mode. The paraxial formulation
given below will not only allow us to quantify the directionality
in terms of a Purcell $\beta$-factor for emission into the desired
mode, but also to show that the optimal phases for the master atom
- ensemble couplings $J_{j}$ are naturally generated by Laguerre-Gauss
laser beams driving the transfer from master atom to atomic ensemble.

In the paraxial approximation the photon wavepacket, propagating dominantly
along a given direction (chosen in the following as the $z$-axis
in Figs.~\ref{fig:chiralQO} and \ref{fig:setup}), can be expanded in the form %
\mbox{%
$\vec{\psi}(\vec{\rho},z,t)=\sum_{n}\psi_{n}(t-z/c)u_{n}\left(\vec{\rho},z\right)\vec{p}\,e^{ik_{0}z}$%
}, where we denote $(\vec{\rho},z)\equiv\vec{r}$. Here $u_{n}\left(\vec{\rho},z\right)$
is a complete set of (scalar) modes solving $\left(\partial_{z}-\frac{i}{2k_{0}}\nabla_{\perp}^{2}\right)u_{n}\left(\vec{\rho},z\right)=0,$
and satisfying for a given~$z$ the orthogonality condition %
\mbox{%
$\int d^{2}{\vec{\rho}}\,u_{n}^{*}\left(\vec{\rho},z\right)u_{m}\left(\vec{\rho},z\right)=\delta_{nm}$%
} \footnote{In practice, the range of $n$ is limited by the condition that the spectral distribution of $u_n(\vec \rho,z)$ takes non-vanishing values only for momenta $|\vec q|\ll k_0$. }. Examples of paraxial modes include the Laguerre-Gauss modes $LG_{p}^{l}(\vec{\rho},z)$
with radial and azimuthal indices $p$ and $l$. The LG modes are
implicitly parametrized by the beam waist $w_{0}$, and the focal
point $z_{0}$, as summarized in Appendix \ref{sec:-Laguerre-Gauss-(LG)}. 

Expanding the field emitted from the antenna into a set of paraxial
modes allows to decompose the spontaneous decay rate $\gamma_{{\rm tot}}$
of the master atom as $\gamma_{{\rm tot}}=\sum_{n}\gamma_{n}+\gamma^{\prime}$.
Here $\gamma_{n}$ is the spontaneous emission rate into the paraxial
mode $u_{n}(\vec{\rho},z)$, while $\gamma^{\prime}$ denotes the
emission into the remaining modes in $4\pi$ solid angle. In Appendix
\ref{sec:Emission-rate-into} we derive 
\begin{equation}
\gamma_{n}=\frac{3\pi\gamma_{e}}{2k_{0}^{2}}\Big|\sum_{i,j}u_{n}^{*}\left(\vec{\rho}_{i},z_{i}\right)e^{-ik_{0}z_{i}}\left(H_{nh}\right)_{i,j}^{-1}J_{j}\Big|^{2}\label{eq:gamman}
\end{equation}
which is essentially the spontaneous emission rate according to Fermi's
golden rule with the emitted field pattern $\vec{\varphi}(\vec{r}\,)$
projected on the paraxial modes $u_{n}(\vec{\rho},z)$. This leads
us to define a Purcell factor 
\begin{equation}
\beta_{n}\equiv\frac{\gamma_{n}}{\gamma_{{\rm tot}}},\label{eq:beta_of_delta}
\end{equation}
as the fraction of the total emission into each paraxial mode $n$
($0\le\beta_{n}\le1$). Our aim is thus to find a set of couplings
$\{J_{j}\}$ which optimizes emission into a given directed mode —
say a target mode $n_{0}$ — ideally with $\beta_{n_{0}}\rightarrow1$.

Purcell factors close to unity for a given mode can be achieved for
off-resonant transfer $\Delta\gg\gamma_{e}$. In this limit we have
\mbox{%
$H_{nh}^{-1}\approx-\Delta^{-1}\mathbb{I}+i\gamma_{e}/\left(2\Delta^{2}\right)\left(\mathbb{I}+\mathbb{G}\right)$%
} up to second order in $1/\Delta$, and dipolar flip-flops in the
atomic ensemble are suppressed as higher order terms in a large detuning
expansion. We then find 
\begin{align}
\gamma_{n} & = \left|g_n\right|^2,\ \ \  \ \ g_n=\frac{\sqrt{3\pi\gamma_{e}}}{\sqrt{2}\Delta k_{0}}\sum_{j}u_{n}^{*}\left(\vec{\rho}_{j},z_{j}\right)e^{-ik_{0}z_{j}}J_{j},\label{eq:gamma_n_large_Delta}\\
\gamma_{{\rm tot}} & =\frac{\gamma_{e}}{\Delta^{2}}\sum_{i,j}J_{i}^{*}\left(\mathbb{I}+{\rm Re}[\mathbb{G}]\right)_{ij}J_{j}.\label{eq:gamma_tot_large_Delta}
\end{align}
From this expression we see that the emission rate $\gamma_{n_{0}}$
to the target mode of interest $n_{0}$ is maximized under the prescription
\begin{equation}
J_{j}\sim e^{ik_{0}z_{j}}u_{n_{0}}\left(\vec{\rho}_{j},z_{j}\right),\label{eq:Jj_choice}
\end{equation}
while other $\gamma_{n\ne n_{0}}$ are strongly suppressed, as a consequence
of the orthogonality condition of the paraxial modes in a discrete
approximation. This is a manifestation of the Huygens-Fresnel principle, where the atomic emission interferes constructively in the desired direction. We emphasize that these couplings are naturally implemented
in the physical setup of Fig.~\ref{fig:chiralQO}(a,c), when the laser driving
the master atom --- ensemble couplings is chosen with the spatial mode
$u_{n_{0}}$. Under the prescription \eqref{eq:Jj_choice}, we obtain for the effective decay rate of Eq.~\eqref{eq:gamma_n_large_Delta} \mbox{$\gamma_{n_0}=3\pi \gamma_e \bar J^2 N_z/(2\Delta^2k_0^2\delta_\perp^2)$}, where $\bar J\equiv \sqrt{\sum_j|J_j|^2}$. As an example, for $\bar J\sim 0.15 \Delta$, $N_z=2$ and $\delta_\perp=0.7\lambda_0$, this corresponds to $\gamma_{n_0}\sim 10^{-2} \gamma_e$, which for $\gamma_e$ in the MHz range represents timescales of the order of $10^{-4}$ s.

We note that in the limit $\Delta\gg\gamma_{e}$ of off-resonant excitation
the atoms representing the quantum antenna are only virtually excited,
i.e. the atomic ensemble acts as a \emph{virtual quantum memory}.
This is in contrast to \emph{real quantum memory} for quantum states
of light in atomic ensembles \cite{Hammerer2010}, where an incident
photon is absorbed and stored in a long-lived spin excitation, and
is read out after some storage time in a Raman process.

In the following section we will present a numerical study for various
antenna geometries, including ensembles as regular atomic arrays,
and for randomly positioned atoms. Analytical results for Purcell
factors can be obtained in the limit of a large number of randomly
distributed atoms, and assuming the choice of phases as given by \eqref{eq:Jj_choice}.
In Appendix~\ref{sec:purcell random} we show that for such a `random'
ensemble with atomic density $n_{a}$ and optical depth $\smash{\mathcal{O}_{d}=3\lambda_{0}^{2}n_{a}L_{z}/(2\pi)}$,
the Purcell factor can be written as $\beta_{n_{0}}=\mathcal{O}_{d}/\left(4+\mathcal{O}_{d}\right)$.
This is consistent with the expression for readout efficiency of ensemble
based atomic quantum memories \cite{PhysRevA.76.033805,Saffman2005}. Thus $\beta_{n_{0}}\rightarrow1$
is achievable in the limit of large optical depth, i.e. large number
of atoms. Remarkably, as shown in the following section, the number
of atoms required can be significantly relaxed for regular arrays
with subwavelength spacing.

\subsection{Numerical Study of Few-Atom Arrays\label{subsec:Few-atom-antenna}}

We now turn to a numerical study for characterizing and optimizing
the geometry of the quantum antenna. We will show in particular that
regular atomic arrays, due to their periodic structure, can significantly
suppress spontaneous emission into non-forward propagating modes and
this will allow us to achieve large Purcell factors even for a few-atom
antenna. In our study below we choose as target mode a Gaussian beam
$u_{n_{0}\equiv(0,0)}(\vec{\rho},z)\equiv LG_{0}^{0}(\vec{\rho},z)$
(for notation see Appendix~\ref{sec:-Laguerre-Gauss-(LG)}) with a
beam waist $w_{0}$ and the antenna located at the focal point $z_{0}=0$
{[}see Fig.~\ref{fig:setup}(b){]}\footnote{Such modes can be further transformed by optical lenses to interface
the master atom with another system of interest, such as an optical
fiber, a nanophotonic device, an extra master atom, etc.}. Furthermore, we assume that phases are chosen as fixed according
to the prescription \eqref{eq:Jj_choice}, i.e. as $J_{j}\sim e^{ik_{0}z_{j}}LG_{0}^{0}(\vec{\rho}_{j},z_{j})$.

Let us consider an antenna with a configuration characterized by a
transverse atom number $N_{\perp}$, a number of layers $N_{z}$,
and thus a total atom number \mbox{$N_{{\rm a}}=N_{\perp}\times N_{\perp}\times N_{z}$}.
In the following we fix the distance between the layers as $\delta_{z}=\lambda_{0}(2N_{z}-1)/(2N_{z})$,
which provides maximum destructive interference and thus suppression
of emission in the backward direction, but we leave open the transverse
distance $\delta_{\perp}$ as a parameter to be varied. The quantity
to be optimized is the Purcell factor for the Gaussian mode. We find
the maximum value of $\beta_{n_{0}}(w_{0},\delta_{\perp})$, denoted
$\beta$, by varying the waist parameter $w_{0}$ and the transverse
spacing $\delta_{\perp}$.

Our results are shown in the inset of Fig.~\ref{fig2}(a), which
shows $\beta$ as a function of~$N_{{\rm a}}$ for \emph{perfectly
regular arrays} with $N_{z}=2,4,8$ in blue, orange, and green lines,
respectively. Remarkably, a large Purcell factor of $\beta\approx0.94$
can be reached with just two layers of $\smash{4 \times 4}$ arrays
of atoms as antenna (blue square marker on the blue curve), and we
see rapid convergence to 1 with increasing atom number. We note that for the \emph{perfect} arrays considiered here, the efficiency decreases with increasing $N_z$ at fixed $N_\perp$, which is a consequence of the divergence of the paraxial mode increasing with the longitudinal antenna size, such that the optimal configuration is $N_z=2$.
The corresponding
waists $w_{0}$ are shown in blue circles in Fig.~\ref{fig2}(b),
as a function of transverse antenna size $L_{\perp}\equiv N_{\perp}\delta_{\perp}$,
for all configurations of $N_{\perp}$ and $N_{z}$ up to $\smash{16 \times 16 \times 8}$.
The dashed blue line indicates the linear dependence of the optimal
mode waists $w_{0}\sim L_{\perp}$ on the transverse antenna size
(as long as $w_{0}\gtrsim\lambda_{0}$). The optimization routine
identifies the largest mode waist supported by the antenna, as the
Purcell factor for regular arrays increases with the growth of the
transverse mode size, as we discuss below. The red dashed curve in
Fig.~\ref{fig2}(b) shows the corresponding opening angle of the
Gaussian mode as given by $\theta=\tan^{-1}[\lambda_{0}/(\pi w_{0})]$.
The optimal interatomic spacings $\delta_{\perp}$ are presented in
Fig.~\ref{fig2}(c) and indicate a slow growth with the increase
of the transverse antenna size. This is a consequence of the fact
that the transverse spatial spectrum of the target mode becomes narrower
for larger antennas and, according to the sampling theorem, the antenna
can properly couple to the mode even with an increasing interatomic
spacing $\delta_{\perp}$.

To compare regular arrays and `random' atomic ensembles, and to reveal
the antenna performance scaling with its size, we find it convenient
to define an effective optical depth for regular arrays as $\mathcal{O}_{d}^{{\rm eff}}=4\beta/(1-\beta)$,
corresponding to the optical depth for a `random' ensemble achieving
the same Purcell factor. This effective optical depth is shown in
Fig.~\ref{fig2}(a) for regular perfect arrays with $N_{z}=2,4,8$
in blue, orange, and green solid lines, respectively. For comparison,
results for an antenna with randomly distributed atoms, with atomic
density equivalent to the one of regular arrays, are shown in black
dotted line.

The performance of the $\smash{4 \times 4}$ bilayer
array mentioned above is highlighted by the optical depth $\mathcal{O}_{d}^{{\rm eff}}\approx70$
(blue square marker). The scaling of the optical depth with the number
of atoms shows a striking difference between regular arrays and `random'
atomic ensembles. This is due to the fact that even though the emission
rate $\gamma_{n_{0}}$ into a target mode given by Eq.~\eqref{eq:gamma_n_large_Delta}
is similar, the total emission rate, which for the optimized couplings
choice reads $\gamma_{{\rm tot}}\approx\gamma_{n_{0}}+\gamma'$, is
defined mainly by the scattering into non-paraxial modes $\gamma'$.
An ensemble of randomly distributed atoms emits almost equally well
into all non-paraxial modes, although the ratio $\gamma'/\gamma_{n_{0}}$
is suppressed by the number of atoms. For regular atomic arrays, however,
the sideward scattering is totally suppressed for target modes with
large transverse extent. In addition, paraxial backward emission is
significantly suppressed by means of the destructive interference
with the proper choice of longitudinal spacing $\delta_{z}$ given
above. This results in Purcell factors for regular arrays, which are
far superior to the one of a `random' ensemble.

In Appendix~\ref{app:Od_lattice} we show that the effective optical
depth for a lattice emitting into a mode with transverse size $w$
grows as $(w/\lambda_{0})^{4}$, which is equivalent to %
\mbox{%
$\mathcal{O}_{d}^{{\rm eff}}\sim(N_{{\rm a}}/N_{z})^{2}$%
} since $w/\lambda_{0}\sim N_{\perp}$. More precisely, for a Gaussian
mode with waist $w_{0}$ focussed inside an antenna of size $L_{\perp}\gg w_{0}$
we have %
\mbox{%
$\mathcal{O}_{d}^{{\rm eff}}\leq8+32(w_{0}^{4}/\sigma^{2})$%
} for a two layer antenna with $\delta_{z}=(3/4)\lambda_{0}$, where
$\sigma=3\lambda_{0}^{2}/(2\pi)$ is the scattering cross section
of a two-level atom, where the upper bound corresponds to the probability
of emitting forwards. This analytical result is shown in black dot-dashed
line in Fig.~\ref{fig2}(a), where we have converted the antenna
size $N_{{\rm a}}$ into a mode waist $w_{0}$ using the linear dependence
of the optimal mode waist on the transverse antenna size {[}blue dashed
line in Fig.~\ref{fig2}(b){]}. This scaling is in contrast to the
`random' ensemble optical depth, which grows like $\mathcal{O}_{d}\sim(N_{{\rm a}})^{1/2}$
for an ensemble geometry optimized for a Gaussian beam~\footnote{The optimal overlap with a Gaussian beam is achieved for an ensemble
of length $2z_{R}\equiv2\pi w_{0}^{2}/\lambda_{0}$ (i.e. twice the
Rayleigh length) and section $S=\pi w_{0}^{2}$. This ensemble contains
$N_{{\rm a}}=2n_{{\rm a}}Sz_{R}=2n_{{\rm a}}S^{2}/\lambda_{0}$ atoms,
such that $\mathcal{O}_{d}\equiv(\sigma/S)N_{{\rm a}}=\sigma(2n_{{\rm a}}N_{{\rm a}}/\lambda_{0})^{1/2}$
with $\sigma$ the resonant scattering cross section of a two-level
atom.}. This scaling is shown in black dotted line in Fig.~\ref{fig2}(a).

\begin{comment}
The analytical calculation is performed for an emission forward into
a solid angle, therefore, it is an upper bound for the numerical estimations
of the effective optical depth corresponding to emission into a Gaussian
mode shown in Fig.~\ref{fig2}(a) with color lines.

The fact that efficiency of the lattice antenna grows simply with
the transverse size $w_{0}$ of the emission mode can be seen in Fig.~\ref{fig2}(b)
where $w_{0}$ (blue dots) of optimal Gaussian modes for all antenna
configurations (up to %
\mbox{%
$16 \times 16 \times 8$%
}) linearly depend on the transverse size $L_{\perp}$ of the antenna.
The optimization routine basically chooses the largest mode waist
supported by the antenna, almost irrespective of the number of atomic
layers $N_{z}$. 
\end{comment}

The effect of imperfections in atomic arrays is shown in Figs.~\ref{fig2}(d,e,f).
Panel (d) illustrates the decrease of Purcell factor $\beta$ due
to a finite percentage of defects (i.e. missing atoms) for an array
with $N_{z}=2,4,8$ layers. Panel (e) shows the effect of temperature,
which is modeled as a classical randomization of atomic positions
normally distributed with variance $\sigma_\text{th}^2\equiv 2n_\text{th}/(m\omega_m)$, where $n_\text{th}$ is the average thermal occupation of motional states, $m$ the atomic mass and $\omega_m$ the vibrational frequency. This is also shown in Fig.~\ref{fig2}(a) with dashed and dot-dashed
lines corresponding to $\sigma_\text{th}=0.01\lambda_{0},\,0.02\lambda_{0}$,
respectively (the color indicates $N_{z}$ as described above). Clearly,
an antenna consisting of larger number of layers $N_{z}$ is less
prone to imperfections since it has more emitters to support the destructive
interference for the backward scattering. Finally, in panel (f) we
study the Purcell factor $\beta$ {[}without the approximations of
Eqs.~\eqref{eq:gamma_n_large_Delta}, \eqref{eq:gamma_tot_large_Delta}{]}
as a function of the detuning $\Delta$ from the resonance for two-level
atoms. One can see that a detuning of the order of the natural linewidth $\gamma_{e}$
is sufficient to reach optimal Purcell factors.

\section{Chiral Photonic quantum Link}
\label{sec:chiralnetworks}

Quantum antennas can be used as a light-matter interface to form \emph{chiral} photonic quantum networks in free-space, with several distant master atoms strongly interacting via a common 1D free space photonic mode [see Fig.~\ref{fig:chiralQO}(b)]. We illustrate the efficiency of this `quantum link' with simulations of deterministic Quantum State Transfer protocols. We then express the dynamics of a more generic photonic network, including possibly many-photon states, in terms of a Quantum Stochastic Schr\"odinger Equation.

\begin{figure}
\includegraphics[width=1\columnwidth]{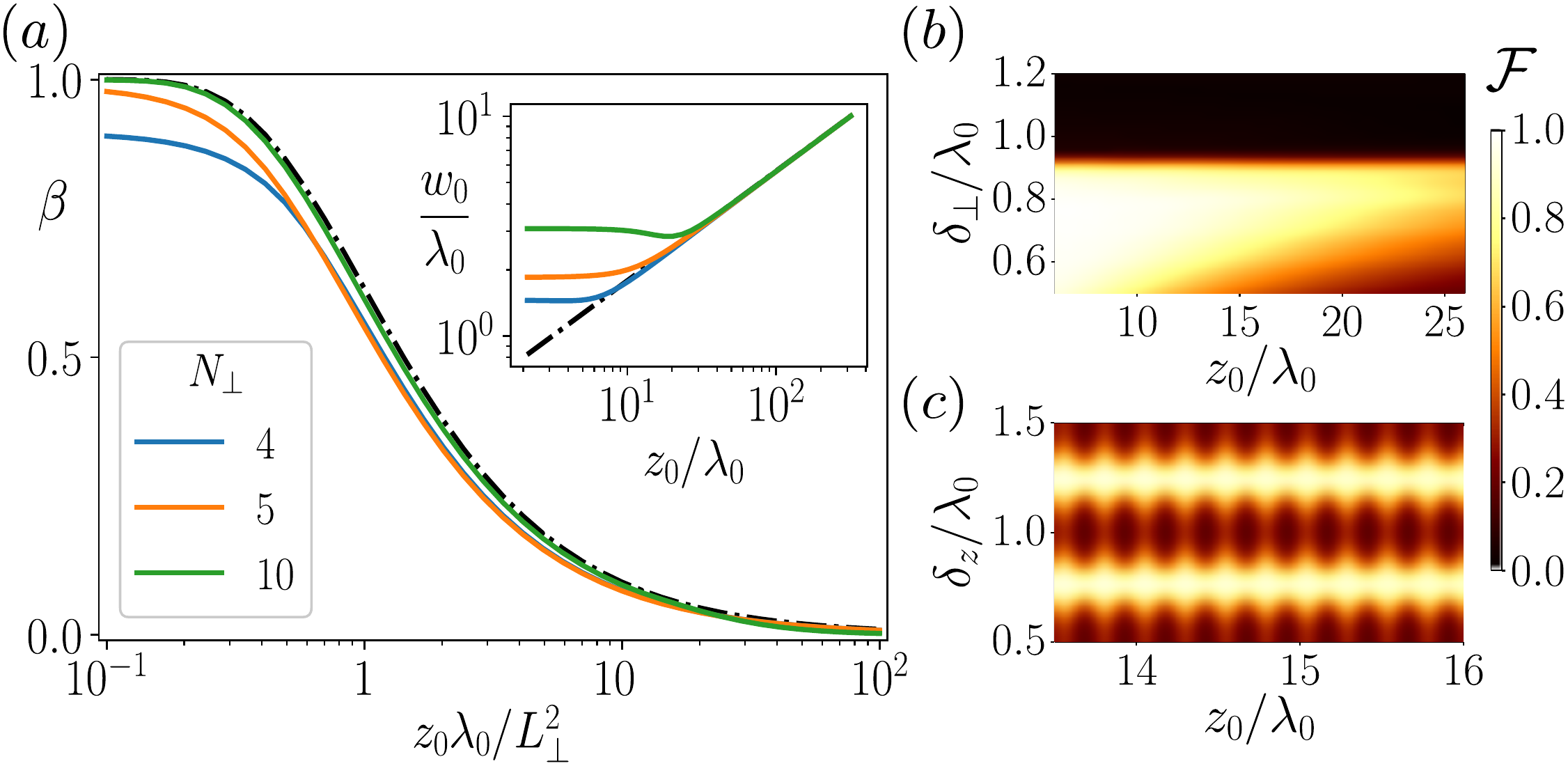}\caption{\emph{Free-space quantum link}. (a)~Purcell factor $\beta$ as a function of the distance $z_0$ from each antenna to the focussing point of the gaussian mode for various $N_\perp$, with $\delta_\perp=0.75\lambda_0$. Inset: corresponding optimal waists. The dotted-dashed curves have analytical expressions (see text).  (b,c)~State transfer fidelity, for $N_\perp=10$ and (b)~$\delta_z=0.75\lambda_0$, (c)~$\delta_\perp=0.8\lambda_0$. }
\label{qsse} 
\end{figure}

\subsection{Chiral Master Equation with Quantum Antennas}\label{sec:QLink}
We consider a minimal network consisting of two nodes separated by a distance $L=2z_0$ [see Fig.~\ref{fig:chiralQO}(b)], where each master atom \mbox{$(a=1,2)$} has a ground state $\ket{G}_a$ and an excited state $\ket{E}_a$, with $s_a^-\equiv\ket{G}_a\!\bra{E}$ and is coupled to an array of $N_{\rm a}$ atoms. We denote the hopping rates as $J_{j,a}$ with $j=1,...,2N_{\rm a}$, where $J_{j,1}$ ($J_{j,2}$) takes non-zero values only for $j\leq N_{\rm a}$ (resp. $j>N_{\rm a}$). We assume that each array acts as a quantum antenna with a common paraxial target mode $n_0$, with the waist of this mode located halfway between the two nodes \footnote{One can alternatively use optical lenses to couple paraxial modes with waists located at the position of each antenna}.  As represented in Fig.~\ref{qsse}(a), a good photon emission and absorption into this mode, characterized by $\beta\approx1$, is realized when $L_\perp\gtrsim\sqrt{\lambda_0 z_0}$, a condition set by the diffraction limit. The dashed-dotted curve represents \mbox{$\int_{|x|,|y|<L_\perp/2} d^2\rho \left|u_{n_0}\left(\vec \rho,z_1\right)\right|^2=\text{erf}\left[\sqrt{L_\perp^2\pi/(4z_0\lambda_0)}\right]^2$}, with $z_1$ the position of the first antenna along $z$, which is the maximum value for $\beta$ achievable with an antenna with transverse surface $L_\perp^2$. This corresponds to the limit $N_\perp\to\infty$, and already for $N_\perp=10$ the curve is almost indistinguishable from this limit.
The corresponding optimal waists are shown in the inset of Fig.~\ref{qsse}(a), where the black curve represents the waist minimizing the mode width at the position of the antenna, and is given by $w_0=\sqrt{z_0L/\pi}$. For discrete arrays with finite $N_\perp$, the optimal waist is given by a trade-off between minimizing this mode width in order to diminish finite-size effects of the layers, and increasing the number of atoms emitting in the mode in order to improve the effective optical depth. This results in a saturation of $w_0\sim L_\perp$ for low values of $z_0$. 

In order to characterize the efficiency of the quantum link between the two qubits, we first express the dynamics in terms of a master equation for the density matrix $\rho$ for the master atoms.  Assuming that (i) the photonic field is initially in the vacuum state, (ii) the couplings are perturbative (i.e.,~$J_{j,a}\ll\Delta$), and (iii) the time-delay in the photon propagation between the two antennas is negligible (i.e.,~Markov approximation), we eliminate antenna atoms and radiation field as an effective reservoir for master atoms, and obtain a chiral master equation analogous to Eq.~\eqref{eq:mastereq1D}, 
where now the effective non-hermitian Hamiltonian reads (see Appendix~\ref{sec:chiralmasterequationderiviation}) 
\begin{equation}
\begin{aligned}
H_{\text{eff}}  =  & \sum_{a=1}^2\left(\epsilon_{a}-i\frac{\gamma_{\text{tot},a}}{2}\right)s_{a}^{+}s_{a}^{-}
\\ &-i\left(\gamma_{L}e^{i\phi_L}s_{1}^{+}s_{2}^{-}+\gamma_{R}e^{i\phi_R}s_{2}^{+}s_{1}^{-}\right).
\label{eq:Hchir}
\end{aligned}
\end{equation}
Here %
\mbox{%
$\epsilon_{a}-i\gamma_{\text{tot},a}/2\equiv-\sum_{i,j}J_{i,a}^{*}(H_{nh}^{-1})_{i,j}J_{j,a}$%
}, where $\epsilon_{a}$ is a frequency shift for each qubit \footnote{$\epsilon_a$ can be cancelled e.g. using additional AC Stark
shifts.}, while $\gamma_{a}$ is their total effective decay rate, similarly to Eq.~\eqref{eq:full_gamma}.
We note that due to the symmetry of the system and of the target mode, we have here $\gamma_{\text{tot},1}=\gamma_{\text{tot},2}\equiv\gamma$. 
On the other hand, the rates of interaction between qubits and the associated phases are given by
\begin{eqnarray}
\label{eq:gammaRphiR}\gamma_{R}e^{i\phi_{R}}= & - & i\sum_{i,j=1}^{2N_{\rm a}}J_{i,2}^{*}(H_{nh}^{-1})_{i,j}J_{j,1},\\
\label{eq:gammaLphiL}\gamma_{L}e^{i\phi_{L}}= & - & i\sum_{i,j=1}^{2N_{\rm a}}J_{i,1}^{*}(H_{nh}^{-1})_{i,j}J_{j,2},
\end{eqnarray}
and the decay to unwanted non-paraxial modes expresses as $\gamma'\equiv\gamma-\gamma_R-\gamma_L$. 
 Finally, the recycling terms express as
\begin{eqnarray}
\mathcal{L}_{\text{eff}}{\rho} & = & \sum_{a=1}^2\gamma_{\text{tot},a}s_{a}^{-}{\rho}s_{a}^{+}+\left(\gamma_{R}e^{i\phi_R}+\gamma_{L}e^{-i\phi_L}\right)s_{1}^{-}{\rho}s_{2}^{+}\nonumber \\
 &  & +\left(\gamma_{R}e^{-i\phi_R}+\gamma_{L}e^{i\phi_L}\right)s_{2}^{-}{\rho}s_{1}^{+},\label{eq:Lchir}
\end{eqnarray}

The rate $\gamma_{R}$ ($\gamma_{L}$) in Eq.~\eqref{eq:Hchir} corresponds to the effective long-range coupling from the first qubit to the second one (second to first, respectively), which in general is not reciprocal (i.e.,~$\gamma_R\neq\gamma_L$). In analogy to Eq.~\eqref{eq:gamma_n_large_Delta}, assuming $k_0z_0\gg 1$, we obtain in the paraxial approximation and to lowest order in $\gamma_e/\Delta$
\begin{equation}
\gamma_{R}e^{i\phi_R}\!=\! \frac{3\pi\gamma_e}{2\Delta^2 k_0^2}\! \sum_{n,i,j}\!e^{ik_0(z_i-z_j)}\!J^*_{i,2}u_n(\vec{\rho}_i,z_i)
u^*_n(\vec{\rho}_j,z_j)J_{j,1},
\end{equation}
while $\gamma_Le^{i\phi_L}$ can be expressed in a similar way by replacing $J_{j,a}\to J_{j,a}^*$. This assumes a decomposition of left-propagating modes on a similar paraxial basis, with the same waist location as for the right-propagating modes. In particular, using Eq.~\eqref{eq:Jj_choice} for the couplings, in the limit $\beta\to 1$ we have the expression \mbox{$\gamma_R\to 3\pi \gamma_e \bar J_1\bar J_2 N_z/(2\Delta^2k_0^2\delta_\perp^2)$}, where $\bar J_a\equiv\sqrt{\sum_i |J_{i,a}|^2}$ denotes the coupling rate between master atom $a$ and its antenna, while $(\gamma_L,\phi_R)\to0$ and $\phi_L\to 4k_0z_0$.
We thus obtain almost ideal unidirectional couplings between qubits, thus forming a {cascaded} quantum system as discussed in Sec.~\ref{sec:chiralQO}. We illustrate this here with the application of deterministic Quantum State Transfer (QST) protocols.

To realize QST, we first remark that the various decay rates in Eq.~\eqref{eq:Hchir} can be taken time-dependent by adding a temporal modulation in the laser-assisted hopping rates $J_{j,a}\to f_a(t) J_{j,a}$, such that \mbox{$\gamma_{\text{tot},a}\to f_a(t)^2 \gamma_{\text{tot},a}$} and $\gamma_{R/L}\to f_1(t)f_2(t)\gamma_{R/L}$. The functions $f_1(t)$ and $f_2(t)$ are chosen such that in the ideal scenario (i.e.,~\mbox{$\gamma_{R}=\gamma$}) the total excitation of the two qubits is conserved~\footnote{Analytical expressions for $f_1(t)$ and $f_2(t)$ can be obtained by requiring the temporal shape of photons emitted by the first array to be symmetric, in which case a solution is $f_2(t)\equiv f_1(-t)$.}.
With imperfect couplings ($\gamma_R<\gamma$), the QST fidelity is given by \mbox{$\mathcal F=\gamma^2_R/\gamma^2$}, assuming $\gamma_L\ll\gamma_R$ (see Appendix~\ref{sec:pulseshapingQST} for details).

Fig.~\ref{fig:chiralQO}(e) represents the range of values for $\mathcal F$ accessible in our model, which are obtained by evolving the dynamics of the master atom density matrix $\rho$ from the chiral master equation, and shows that the typical achievable inter-array separations $2z_0$ grows linearly with the surface $L_\perp^2$. The dashed curve represents a numerical estimation for the maximum fidelity achievable in the paraxial limit, and corresponds to the limit of $N_\perp\to\infty$. In the inset, we show that the saturation value decreases like $1/N_\perp^4$, which is equivalent to the scaling of $\mathcal O_d^\text{eff}\sim(N_{\rm a}/N_z)^2$ as discussed in Sec.~\ref{subsec:Few-atom-antenna}. As an example, with $N_\perp=20$ we obtain $\mathcal F \approx 0.88$ for $2z_0=150\lambda_0$, demonstrating the efficiency of atomic arrays for building optical interconnects with mesoscopic distances. For small $z_0$, Fig.~\ref{qsse}(b) shows that $\delta_\perp$ can take a broad range of values, as long as $\delta_\perp\lesssim0.9\lambda_0$. As $z_0$ increases, this range diminishes as we need a larger surface $L_\perp^2\sim z_0 \lambda_0$. Conversely, by varying $\delta_z$, a spatial periodicity of $\lambda_0/2$ appears when $\delta_z\neq\lambda_0(1\pm1/4)$ [see Fig.~\ref{qsse}(c)], arising from the fact that $\gamma_L$ is not properly cancelled, which induces a back-action on the first qubit depending in general on the propagation phase as $\phi_R+\phi_L= 4k_0z_0$. 

Finally, we note that for fidelities $\mathcal{F}$ close to $1$, several strategies for quantum error correction can be applied to our situation to further improve the fidelity. This can be realized by coupling several qubits to each atomic array, rather than a single one, to implement redundant
qubit codes correcting for the photon losses arising from $\beta<1$. For instance, following a protocol described in Ref.~\cite{Vermersch2017}, the qubit state can first be redundantly encoded in an atomic ensemble, using entangled states with multiple atomic excitations for the logical qubit, such as cat or binomial states~\cite{Michael2016}. Coupling the ensemble to the atomic array will produce a propagating quantum error correcting photonic code rather than a single photon, which can then be transferred to a second distant ensemble, using the same protocol as described above. Provided the probability of error is small enough, single photon losses can be detected and corrected in the second ensemble. On the other hand, when the fidelity $\mathcal F$ is too low for error correction, probabilistic protocols become advantageous, and atomic arrays can be used to achieve high repetition rates.

\subsection{Quantum Stochastic Schr\"odinger Equation Formulation}\label{sec:QSSE}

In Sec.~\ref{sec:Model} we studied the spontaneous emission process of a single photon from a single master atom, which we described using a Wigner-Weisskopf ansatz. In Sec.~\ref{sec:QLink} we then considered a system of two nodes, whose generic dynamics was provided by a master equation, obtained under the assumptions that the radiation was initially in the vacuum state, and that time-delays in the photon propagation between the nodes was negligible (Markov approximation).
In the following we extend these formalisms to account for such possible delays, and to allow for the description of any initial photonic field state.

\subsubsection{Single emitter}
We start with the description of a single node of master atom and quantum antenna.
Beyond the Wigner-Weisskopf treatment presented in Sec.~\ref{sec:Model}, the dynamics, including possibly many-photon states, is conveniently formulated in the framework of quantum stochastic calculus, in terms of a Quantum Stochastic Schr\"odinger Equation (QSSE) \cite{gardiner2015}. Our description is obtained in the limit $\Delta\gg\gamma_e$ by adiabatically eliminating excitations in the antenna in an Holstein-Primakoff approximation. For details on the derivation of the QSSE and formal definition of the field modes we refer the reader to Appendix~\ref{app:qsse}. We obtain
\begin{equation}\label{eq:qssedef}
i\frac d {dt}\ket{\Psi(t)}=\left[H_\text{sys}+V(t)\right]\ket{\Psi(t)},
\end{equation}
describing the dynamics of a pure state $\ket{\Psi(t)}$ including master atom and photonic field,
where the interaction Hamiltonian expresses as
\begin{equation}\label{eq:QSSE_Vdef}
V(t)=i\Big[\sum_n\left(g^R_n {b^R_n}^\dagger(t)+g^L_n {b^L_n}^\dagger(t)\right)
+g'{b'}^\dagger(t)\Big] s^-+\text{h.c.} 
\end{equation} 
Here $b^{R/L}_n(t)$ represent quantum noise annihilation operators for photons in the paraxial 1D mode $n$ propagating in the right/left direction, and interacting with the master atom at time $t$, while $b'(t)$ corresponds to unwanted modes propagating in 3D, satisfying \mbox{$[b^{p}(t),{b_n^q}^\dagger(t')]=\delta_{p,q}\delta(t-t')$} with $p,q\in\{R,L,'\}$. 

The effective coupling of the master atom to right-propagating modes $g_n^R$ expresses as $g_n$ in Eq.~\eqref{eq:gamma_n_large_Delta}, while $g_n^L$ can be expressed as in Eq.~\eqref{eq:gamma_n_large_Delta} by replacing $u_{n}\left(\vec{\rho}_{j},z_{j}\right)e^{ik_{0}z_{j}} \to u_{n}^{*}\left(\vec{\rho}_{j},z_{j}\right)e^{-ik_{0}z_{j}}$.  The coupling $g'$ on the other hand expresses from Eq.~\eqref{eq:gamma_tot_large_Delta} as \mbox{$g'=\sqrt{\gamma_\text{tot}-\sum_n(|g_n^R|^2+|g_n^L|^2)}$}. For generality sake we also added a term $H_\text{sys}$, accounting for eventual additional operations on the master atom which needs not conserve the number of excitations, e.g. for an external coherent drive \mbox{$H_\text{sys}=\Omega_\text{R}(s^-+s^+)-\Delta_\text{R} s^+s^-$} with Rabi frequency $\Omega_\text{R}$ and detuning $\Delta_\text{R}$ \footnote{The QSSE can be interpreted according to quantum Stratonovich calculus and integrated to obtain a master equation for the master atom depending on the state of the input fields.}.
The dynamics generated by Eq.~\eqref{eq:QSSE_Vdef} is in exact analogy to that of a qubit with {chiral} coupling ($g^L_n\neq g^R_n$) to a multimode 1D waveguide, where each $n$ corresponds to an orthogonal degenerate waveguide mode, achieving ideal coupling in the limit $g'\to0$. 

\subsubsection{Two emitters}

We now consider the case of two nodes as studied in Sec.~\ref{sec:QLink}.
The interaction Hamitonian in Eq.~\eqref{eq:QSSE_Vdef}  expresses here as $V(t)=V_1(t)+V_2(t)$, where (see Appendix~\ref{app:qsse})
\begin{eqnarray}
V_a(t)=i\Big[g_a'{b_a'}^\dagger(t)+\sum_n &&\Big(g^R_{n,a}e^{-i\omega_0\tau_a}{b^R_n}^\dagger(t-\tau_a)\label{eq:QSSE_V2def}
\\+&&g^L_{n,a}e^{i\omega_0\tau_a}{b^L_n}^\dagger(t+\tau_a)\Big)\Big]s_a^- +\text{h.c.},\nonumber
\end{eqnarray}
with $\tau_1=0$, and $\tau_2\equiv\tau= 2z_0/c$ the time-delay in the propagation of a photon between the two nodes. Here ${b_{1,2}'}(t)$ denote annihilation operators for photons in non-paraxial modes, which we assume independent with $[{b_1'}(t),{b_2'}^\dagger(t')]=0$, and $g'_{1,2}$ denote the effective coupling rates of the master atoms to these modes. On the other hand $b^{R}_n(t)$ represent bosonic operators for a continuous string of harmonic oscillators interacting consecutively with the first and second master atom, while $b^{L}_n(t)$ represent harmonic oscillators interacting consecutively with the second and first atom. 

%In general, the time-delay $\tau$ in the label of field operators of Eq.~\eqref{eq:QSSE_V2def} will yield a non-markovian dynamics, that is, with the field effectively retaining a memory of the atomic state at previous times, thereby violating an assumption for the derivation of an atomic master equation for the density matrix $\rho$ of master atoms. 

In some cases, the time-delay $\tau$ in the label of field operators of Eq.~\eqref{eq:QSSE_V2def} can be formally set to \mbox{$0^+$}, such that the QSSE can be integrated. This is the case when the time-delay is shorter than the typical timescale of the atomic dynamics (Markov approximation, i.e.~\mbox{$|g'_a|^2+\sum_n(|g^R_{n,a}|^2+|g^L_{n,a}|^2)\ll1/\tau$}), or for purely cascaded systems (i.e.,~$\sum_n|g^L_{n,a}|^2\ll \sum_n|g^R_{n,a}|^2$),  where photons flow from the first to second node without back-action. In the case of vacuum initial state for the photonic field we obtain the chiral master equation of Sec.~\ref{sec:QLink}, in the paraxial approximation, where we identify \mbox{$\gamma_Re^{i\phi_R}=\sum_n g^R_{n,1} \left(g^R_{n,2}\right)^*$} and \mbox{$\gamma_L e^{i\phi_L}=\sum_n g^L_{n,2}\left(g^L_{n,1}\right)^*$}.  When the delay cannot be neglected however, numerical techniques can be used to solve the QSSE, such as matrix-product-state methods.

\section{Atomic Implementation \label{sec:Atomic-Implementations}}

Quantum antennas can be realized in various microscopic systems. The
basic requirements for the physical realization of the model of the
previous section, as master atom (qubit) coupled to a quantum antenna,
are the following. (i) Excitations must be transferred \emph{coherently}
from master atom (qubit) to the antenna atoms. For a quantum antenna
built as large atomic arrays this requires long-range couplings. (ii)
The spatial distribution of the corresponding couplings $J_{i}$,
in particular the required phases for directional emissions, can be
\emph{engineered}, for instance using laser-assisted processes {[}see
Eq.~\eqref{eq:Jj_choice}{]}. (iii) Antenna atoms can emit photons
via an optical dipole transition.

\begin{figure}
\includegraphics[width=1\columnwidth]{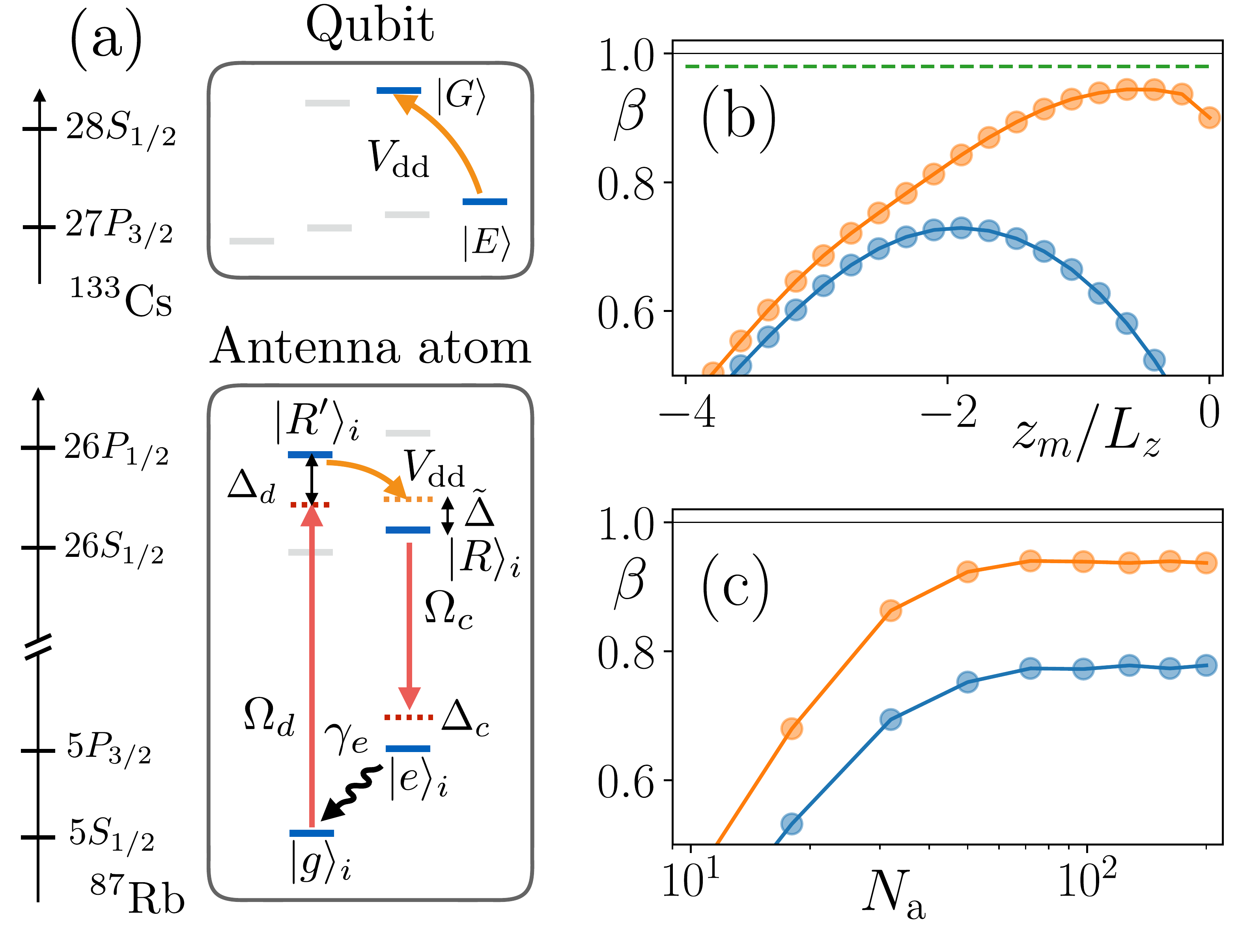}\caption{\emph{Rydberg implementation}. (a) Atomic level schemes, with the
relevant states for our implementation in blue. (b,c) Probability
of emission in a Gaussian mode (b) as a function of the distance $z_{m}$
between master atom and antenna {[}dashed line: value for two-level
model, see Fig.~\ref{fig2}(a){]}, and (c) as a function of the number
of atoms for the optimal $z_{m}$, using dressing laser with a LG
profile (blue), or with an optimized spatial distribution of $\Omega_{d}(\vec{r}_{i})$
(orange). We consider an ensemble of $10\times10\times2$ atoms with
$\delta_{\perp}=0.7\lambda_{0}$, $\delta_{z}=1.25\lambda_{0}$, $\Omega_{c}=2\pi\times2.5\,\text{MHz}$,
$|\Omega_{d}/\Delta_{d}|\leq0.02$.}
\label{impln-1} 
\end{figure}

Here we present an implementation
with neutral atoms employing laser-assisted Rydberg interactions.
This builds on the recent experimental progress in loading atoms in
regular optical lattices, using e.g. optical trapping techniques,
and the possibility to laser excite atoms to Rydberg states to induce
and control long-range dipolar interactions. 
We remark that our model can also be realized in various other atomic
physics setups. For very small antenna sizes, e.g. the minimal antenna
$\smash{2 \times 2 \times 2}$, we can use the physics of atomic
Hubbard models (including synthetic gauge fields~\cite{Goldman2016})
to implement the model of Sec.~\ref{sec:Model}. In addition, for
neutral atoms~\cite{Lahaye2009,Browaeys2016,Glaetzle2017} and molecules~\cite{Micheli2006}
long range coupling are available as magnetic and electric dipolar
interactions. Finally, these ideas can also be translated to a solid-state
context, using quantum dots~\cite{Lodahl2015} or NV centers~\cite{Cai2013}.
This also includes interfacing superconducting qubits~\cite{Houck2012}
stored in strip line cavities with bilayer atomic ensembles acting
as quantum antenna.

The atomic level structure we have in mind is shown in Fig.~\ref{impln-1}(a).
For concreteness, we consider optically trapped atoms in a bilayer
($\smash{N_{z}=2}$) configuration, where the master atom is a $^{133}\text{Cs}$
atom and the antenna is made of $^{87}\text{Rb}$ atoms \footnote{The scheme can be adapted for systems with single atomic species.}. The state
of the master atom is encoded in two Rydberg states $\ket{G}=\ket{28S_{1/2},m_{j}=\frac{1}{2}}$
and $\ket{E}=\ket{27P_{3/2},m_{j}=\frac{3}{2}}$, with microwave transition
frequency $\omega_{\mathrm{Cs}}$ (the quantization axis is set by
an external magnetic field along $z$). Antenna atoms can be excited
to four electronic levels, including two Rydberg states $\ket{R'}_{i}=\ket{26P_{1/2},m_{j}=-\frac{1}{2}}$
and $\ket{R}_{i}=\ket{26S_{1/2},m_{j}=\frac{1}{2}}$, with transition
frequency $\omega_{\mathrm{Rb}}$~\footnote{The nuclear spin $m_{I}=3/2$ can be considered as spectator due to
the small hyperfine interactions between Rydberg levels.}. Our particular choice of Rydberg states is motivated by the small
energy difference $\tilde{\Delta}\equiv\omega_{\mathrm{Cs}}-\omega_{\mathrm{Rb}}=2\pi\times1.74(2)$
GHz between the two Rydberg transitions due to a Förster resonance~\cite{Walker2008}.
Finally, we choose two hyperfine stretched states, a ground state
$\ket{g}_{i}=\ket{5S_{1/2},F=2,m_{F}=2}$ and an excited state $\ket{e}_{i}=\ket{5P_{3/2},F=3,m_{F}=3}$,
in order to generate optical photons with $\lambda_{0}=780$ nm (D2-line).
In our model, we operate in the frozen gas regime \cite{Browaeys2016},
where the motion of the atoms can be neglected for the timescales
associated with our model.

While the complete atomic physics details are presented in Appendix~\ref{sec:Details-of-Rydberg},
we describe here the main elements allowing this setup to behave as
quantum antenna. First, the antenna atoms are subject to a laser beam
coupling off-resonantly, with spatial Rabi frequencies $\Omega_{d}(\vec{r}_{i})$,
and detuning $\Delta_{d}$, $\ket{g}_{i}$ to $\ket{R'}_{i}$. In
the dressing regime, $\Omega_{d}(\vec{r}_{i}),V_{\mathrm{dd}}\ll\Delta_{d}$,
where $V_{\mathrm{dd}}$ is the dipole-dipole coupling between Rydberg
states (see Fig.~\ref{impln-1}), we obtain an effective coherent
`flip-flop' interaction $\smash{\ket{E}\ket{g}_{i}\to\ket{G}\ket{R}_{i}}$
between master atom and antenna, which can be controlled externally
via $\Omega_{d}(\vec{r}_{i})$, i.e we can use the dressing laser
to write the required phases on the antenna atoms. Second, emission
of optical photons from $\ket{R_{i}}$ is assisted by a control laser
with Rabi frequencies $\Omega_{c}(\vec{r}_{i})$, and detuning $\Delta_{c}$.

To show that good directionality can be achieved with realistic configurations,
we now present numerical simulations showing the Purcell factor $\beta$
for emission of a single photon to a Gaussian mode, which include
unwanted dipole-dipole couplings between antenna atoms, and finite
Rydberg states lifetimes. To complement this analysis, we also present
in Appendix~\ref{sec:Details-of-Rydberg} a mapping to Eq.~\eqref{eq:S_P},
which is valid under the condition of the electromagnetically induced
transparency \textbf{ $\Omega_{c}(\vec{r_{i}})\gg J_{i},\gamma_{r}$},
and $\Delta_{c}=0$, with $\gamma_{r}$ being the Rydberg decay rate.
In Fig.~\ref{impln-1} (b,c) we show the Purcell factor $\beta$
for the emission to Gaussian modes. Panel (b) shows that an almost
perfect fidelity of coupling to a Gaussian target mode can be reached
at a certain optimal qubit-antenna distance $z_{m}$. The latter results
from a tradeoff between an exaggerated inhomogeneity of the dipole-dipole
couplings $J_{i}$ at small $z_{m}$ and the predominance of unwanted
losses from the Rydberg states at large $z_{m}$. In panel (c) we
show that the effect of inhomogeneity can be significantly mitigated
by using an optimized spatial distribution of Rabi frequencies $\Omega_{d}(\vec{r}_{i})$
(instead of LG mode). This shows that the atomic antenna based on
Rydberg atoms can be realized with state-of-art technology and with
realistic parameters.

\section{Outlook}

In the present work we propose a scheme for implementing high-efficiency quantum links in free-space, using phased atomic arrays as chiral atom-light optical interface. Our setup realizes the paradigmatic model of chiral quantum optics, with distant atoms interacting via emission and absorption of unidirectional photons.  This allows for the implementation of modular architectures for quantum information processing in free space, without use of dielectric nanostructures or cavities.
In particular, we show that strong connectivity can be achieved even for moderate antenna sizes, allowing for high-fidelity state transfer between atomic qubits.

Quantum antennas as a free-space quantum light matter interface can be extended to incorporate `built-in' modules for quantum information processing. 
Beyond the case of quantum state transfer with single photons discussed here, quantum antennas can be used for instance to generate, emit and absorb photonic states with can error correct for single photon losses~\cite{Michael2016}, and thus increase the free-space link connectivity.

%Beyond the dispersive regime considered here, our setup can be studied with the arrays resonantly coupled to the atoms;
%for instance, the inherent atomic non-linearity of the antenna could be exploited in order to transfer complex multiphoton states, such as quantum error correcting photonic codes~\cite{Michael2016}.

\emph{Note added ---} Since our first submission arXiv:1802.05592v1, the Rydberg coupling between a master atom and an atomic ensemble was also discussed for directional single-photon sources in arXiv:1806.07094v1.  
  
\begin{acknowledgments}
We thank M.~Saffman and F.~Robicheaux for discussions on the Rydberg implementation and valuable feedback on the manuscript, and we acknowledge discussions with R.~Blatt, T.~Monz, M.~Lukin and J.~I.~Cirac. The parameters for the Rydberg simulations were obtained
using the ARC library~\cite{Sibalic2017}. This work was supported
by the Army Research Laboratory Center for Distributed Quantum Information
via the project SciNet, the ERC Synergy Grant UQUAM and the SFB FoQuS
(FWF Project No. F4016-N23). 
\end{acknowledgments}

\bibliography{antenna}

\appendix
%dummy comment inserted by tex2lyx to ensure that this paragraph is not empty

\section{Properties of Laguerre-Gauss Beams}

\label{sec:-Laguerre-Gauss-(LG)} Here we summarize notation, properties
and parametrization of Laguerre-Gauss (LG) modes as used repeatedly
in the main text. LG modes are particular solutions of the paraxial
equation $\left(\partial_{z}-\frac{i}{2k_{0}}\nabla_{\perp}^{2}\right)LG_{p}^{l}\left(\vec{\rho},z\right)=0,$
and are defined, for a given optical wavelength $\lambda_{0}$, by
the mode \emph{waist} $w_{0}$ {[}see Fig.~\ref{fig:setup}{]} as
\begin{equation}
\begin{aligned}LG_{p}^{l}\left({\vec{\rho}},z\right)= & \sqrt{\frac{2p!}{\pi(p+|l|)!}}\frac{1}{w(z)}\left(\frac{\rho\sqrt{2}}{w(z)}\right)^{|l|}e^{-\frac{\rho^{2}}{w^{2}(z)}}\\
 & L_{p}^{|l|}\left(\frac{2\rho^{2}}{w^{2}(z)}\right)e^{i\frac{k_{0}\rho^{2}}{2R(z)}+il\phi-i(2p+|l|+1)\xi(z)},
\end{aligned}
\end{equation}
where $\rho\equiv|{\vec{\rho}}|$ and $\phi=\text{atan}(y/x)$, which
defines a basis with $\int d^{2}{\rho}u_{p,l}\left({\vec{\rho}},z\right)\big(u_{p',l'}\left({\vec{\rho}},z\right)\big)^{*}=\delta_{l,l^{\prime}}\delta_{p,p^{\prime}}.$
For simplicity we assume here that the origin for $z$ is located
at the waist. Here $L_{p}^{|l|}$ denotes the generalized Laguerre
polynomials, where the radial index $p\geq0$ and the azimuthal index
$l$ are integers. The curvature radius is defined as %
\mbox{%
$R\left(z\right)=z+z_{R}^{2}/z$%
}, and the Gouy phase as %
\mbox{%
$\xi(z)=\text{tan}^{-1}(z/z_{R})$%
}, with $z_{R}$ the Rayleigh length %
\mbox{%
$z_{R}\equiv\pi w_{0}^{2}/\lambda_{0}$%
}. The mode width at position $z$ is expressed as %
\mbox{%
$w(z)\equiv w_{0}\sqrt{1+{z^{2}}/{z_{R}^{2}}}$%
}.

\section{Emission Rate into Paraxial Mode\label{sec:Emission-rate-into}}

In this section we derive Eq.~\eqref{eq:gamman} of Sec.~\ref{subsec:Parxial}
for the emission rates $\gamma_{n}$ to a set of orthogonal paraxial
modes $u_{n}\left(\vec{\rho},z\right)$. The photon flux into these
modes is defined through the overlap of the emitted field with the
corresponding modes, in a plane transverse to the propagation axis
$z$, and located at $z_{p}\gg\lambda_{0}$ to neglect the near-field
effects. This allows us to identify 
$\gamma_{n}=c\left|\int d^{2}\vec{\rho}\,u_{n}^{*}(\vec{\rho},z_{p})\vec{p}^{\,*}\vec{\varphi}(\vec{r}\,)\right|^{2}.$
Taking $z_{p}$ far enough from the antenna, only the paraxial part
of $\varphi(\vec{r}\,)$ will contribute, and we can replace the Green's
function in Eq.~\eqref{eq:phi} by its paraxial counterpart as 
\begin{align}
\vec{p}^{\,*}\hat{G}(\vec{r}-\vec{r}_{i})\vec{p}\to & 3\pi k_{0}^{-2}G_{\text{par}}(\vec{\rho}-\vec{\rho}_{i},z-z_{i})\nonumber \\
= & 3\pi k_{0}^{-2}\sum_{n}e^{ik_{0}(z-z_{i})}u_{n}(\vec{\rho},z)u_{n}^{*}(\vec{\rho}_{i},z_{i}).\label{eq:Gpardecomp-1}
\end{align}
Here we have defined %
\mbox{%
$G_{\text{par}}(\vec{\rho},z)={k_{0}}e^{ik_{0}[z+|\vec{\rho}|^{2}/(2z)]}/{(2i\pi z)}$%
} with the property 
{$\int d^{2}\vec{\rho}\,u_{n}^{*}(\vec{\rho},z_{p})G_{\text{par}}(\vec{\rho}-\vec{\rho}_{i},z_{p}-z_{i})=u_{n}^{*}(\vec{\rho}_{i},z_{i}).$}
This gives Eq.~\eqref{eq:gamman} of the main text.

\section{Purcell Factor for Atomic Ensemble with Random Atomic Positions}

\label{sec:purcell random} In this Section we derive analytically
an expression of Purcell factors for an ensemble of $N_{a}$ atoms,
as quoted in Sec.~\ref{subsec:Parxial}, with atoms randomly distributed
in a box with volume $V=L_{\perp}^{2}\times L_{z}$. We start from
Eqs.~\eqref{eq:gamma_n_large_Delta} and \eqref{eq:gamma_tot_large_Delta}
in the limit of large detuning $\Delta\rightarrow\infty$ and write
\begin{equation}
\beta_{n}=\frac{1}{\vec{J}^{\dagger}\left[\mathbb{I}+\text{Re}(\mathbb{G})\right]\vec{J}}\frac{3\pi}{2k_{0}^{2}}\left|\sum_{i}u_{n}^{*}\left(\vec{\rho_{i}},z_{i}\right)e^{-ik_{0}z_{i}}J_{i}\right|^{2},\label{eq:beta_n}
\end{equation}
where, for convenience, we use the vector notation \mbox{$\vec{J}\equiv\left\{ J_{1},\ldots J_{N}\right\} $}.
From Eq.~\eqref{eq:Gpardecomp-1} we have for the term in denominator
\mbox{$\vec{J}^{\dagger}\text{Re}(\mathbb{G})\vec{J}\approx\frac{3\pi}{2k_{0}^{2}}\sum_{n}\left|\sum_{i}u_{n}^{*}\left(\vec{\rho}_{i},z_{i}\right)e^{-ik_{0}z_{i}}J_{i}\right|^{2}$}
Substituting this expression in Eq. (\ref{eq:beta_n}) we get
\begin{align}
\beta_{n} & =\frac{\frac{\sigma}{4}\left|\sum_{i}u_{n}^{*}\left(\vec{\rho_{i}},z_{i}\right)e^{-ik_{0}z_{i}}J_{i}\right|^{2}}{\vec{J}^{\dagger}\vec{J}+\frac{\sigma}{4}\sum_{m}\left|\sum_{i}u_{m}^{*}\left(\vec{\rho}_{i},z_{i}\right)e^{-ik_{0}z_{i}}J_{i}\right|^{2}},\label{eq:beta_n1}
\end{align}
where we defined the single atom resonant scattering cross-section
as $\sigma\equiv{3\lambda_{0}^{2}}/({2\pi})={6\pi}/{k_{0}^{2}}$.
Let us assume now that the coefficients $J_{i}$ are chosen in order
to maximize the emission to the mode $u_{n_{0}}$ as %
\mbox{%
$J_{i}\sim e^{ik_{0}z_{i}}u_{n_{0}}\left(\rho_{i},z_{i}\right)$%
}. Assuming the transverse size of the atomic cloud is larger than
the mode waist, i.e. $w_{0}\lesssim L_{\perp}$, and transforming
the sum in Eq.~\eqref{eq:beta_n1} to an integral $\sum_{i}\rightarrow n_{{\rm a}}\int_{V}d^{3}r$,
we get
\begin{align*}
\beta_{n_{0}} & =\frac{\frac{\sigma}{4}\left|n_{\text{a}}L_{z}\right|^{2}}{n_{\text{a}}L_{z}+\frac{\sigma}{4}\left|n_{\text{a}}L_{z}\right|^{2}}\equiv\frac{{\cal O}_{d}}{4+{\cal O}_{d}},
\end{align*}
with the optical depth defined as ${\cal O}_{d}\equiv\sigma n_{\text{a}}L_{z}$.

\section{Optical Depth of a 3D Lattice}

\label{app:Od_lattice}

In this Section we derive an analytical expression for the Purcell
factor and the effective optical depth for a two layer phased array
of emitters, as quoted in Section~\ref{subsec:Few-atom-antenna}.
Here we consider arrays infinite in the transverse directions, and
we neglect polarization effects. The lattice spacings in the longitudinal
and transverse directions are $\delta_{z}$ and $\delta_{\perp}$,
respectively. The array is prepared to emit a photon unidirectionally
with wavevector $k_{0}$ into a mode with transverse spatial distribution
$f(\vec{\rho}\,)$ of a large width $w\gg\lambda_{0}$ with $\lambda_{0}$
the photon wavelength. Therefore, the mode transverse spatial spectrum
$F_{\perp}(\vec{q}\,)=\int f(\vec{\rho}\,)e^{-i\vec{q}\vec{\rho}}d\vec{\rho}$
has a narrow width $q^{{\rm max}}\sim1/w\ll k_{0}$.

The unnormalized probability amplitude $\varphi(\vec{q},k_{z})$ to
emit a photon into a plane wave with wavevector \mbox{$\vec{k}=\{\vec{q},k_{z}\}$}
can be found as the limit $|\vec{r}\,|\to\infty$ of the Eq.~\eqref{eq:phi}
in the main text, for $\vec{r}=|\vec{r}\,|(\vec{k}/|\vec{k}\,|$).
Here we neglect the polarization part of the Green's function and
consider the limit of large detunings $\Delta\gg\gamma_{e}$. For
a two layer array of emitters (located at $z=\pm\delta_{z}/2$) with
the phases fixed, according to the prescription~\eqref{eq:Jj_choice}
(in the main text), to emit light into the transversally wide mode,
i.e. $J_{j}=e^{ik_{0}z_{j}}f(\vec{\rho}_{j})$, the probability amplitude
$\varphi(\vec{q},k_{z})$ reads 
\begin{align*}
\varphi(\vec{q},k_{z}) & =\sum_{j}J_{j}e^{-i\vec{q}\vec{\rho_{j}}-ik_{z}z_{j}}\\
 & \sim\left[e^{-i(k_{0}-k_{z})\delta_{z}/2}+e^{i(k_{0}-k_{z})\delta_{z}/2}\right]F_{\perp}(\vec{q\,}).
\end{align*}
Here the approximation of the sum with the continuous function $F_{\perp}(\vec{q\,})$
becomes exact for a lattice spacing $\smash{\delta_{\perp}<(2q^{{\rm max}})^{-1}}$,
according to the sampling theorem. The ensemble does not emit in the
transverse direction, as the spectrum $F_{\perp}(\vec{q}\,)$ goes
to zero for $q_{x,y}>q^{{\rm max}}$. Thus the photon can be emitted
into paraxial forward and backward modes only.

In order to define the Purcell factor we need to find the corresponding
amplitudes to emit the photon forward and backward. First, we consider
emission into the paraxial backward modes with $k_{z}^{\leftarrow}\approx-k_{0}+(q_{x}^{2}+q_{y}^{2})/(2k_{0})$.
The longitudinal spacing $\delta_{z}$ is chosen to suppress the exact
backward scattering (given by the plane wave with $k_{z}=-k_{0}$,
$q_{x,y}=0$), i.e. $\delta_{z}=\lambda_{0}(2N_{z}-1)/(2N_{z})$.
The backward scattering amplitude for the interlayer spacing $\delta_{z}=(3/4)\lambda_{0}$
reads 
\begin{align*}
\varphi_{{\rm \leftarrow}}(\vec{q}\,) & =\left[e^{-i(k_{0}-k_{z}^{\leftarrow})\delta_{z}/2}+e^{i(k_{0}-k_{z}^{\leftarrow})\delta_{z}/2}\right]F_{\perp}(\vec{q\,})\\
 & \approx-2\sin\left[\frac{3\pi}{8}\frac{(q_{x}^{2}+q_{y}^{2})}{k_{0}^{2}}\right]F_{\perp}(\vec{q\,})
 \approx\frac{q_{x}^{2}+q_{y}^{2}}{k_{0}^{2}}F_{\perp}(\vec{q\,}).
\end{align*}
The probability to emit light backward is proportional to $\int d\vec{q}\left|\varphi_{{\rm \leftarrow}}(\vec{q})\right|^{2}$,
and is given by 
\begin{align*}
P_{{\rm \leftarrow}} & \sim\int\frac{dq_{x}}{k_{0}}\frac{dq_{y}}{k_{0}}\left|\frac{q_{x}^{2}+q_{y}^{2}}{k_{0}^{2}}F_{\perp}(q_{x},q_{y})\right|^{2}\\
 & \approx\frac{1}{2}\int_{0}^{q^{{\rm max}}}\frac{dq_{r}^{2}}{k_{0}^{2}}\left(\frac{q_{r}^{2}}{k_{0}^{2}}\right)^{2}
  =\frac{1}{6}\left(\frac{q^{{\rm max}}}{k_{0}}\right)^{6}\sim\left(\frac{\lambda_{0}}{w}\right)^{6}.
\end{align*}
Here we approximated the spectrum function as a constant for $q_{x}^{2}+q_{y}^{2}\leq q_{{\rm max}}^{2}$
and zero otherwise. On the other hand, the forward scattering amplitude
has a leading term of order $1$, and the probability to emit forward
reads 
\begin{align*}
P_{{\rm \rightarrow}} & \sim\int\frac{dq_{x}}{k_{0}}\frac{dq_{y}}{k_{0}}\left|F_{\perp}(q)\right|^{2}\\
 & \approx\frac{1}{2}\int_{0}^{q^{{\rm max}}}\frac{dq_{r}^{2}}{k_{0}^{2}}1
  =\frac{1}{2}\left(\frac{q^{{\rm max}}}{k_{0}}\right)^{2}\sim\left(\frac{\lambda_{0}}{w}\right)^{2}.
\end{align*}
This allows us to read off the effective optical depth for a
two infinite layers emitting into a transversally confined mode as
$\mathcal{O}_{d}^{{\rm eff}}=\frac{4\bar\beta}{1-\bar\beta}=\frac{4P_{{\rm \rightarrow}}}{P_{{\rm \leftarrow}}}\sim\left(\frac{w}{\lambda_{0}}\right)^{4},$
where $\bar\beta=P_{{\rm \rightarrow}}/(P_{{\rm \rightarrow}}+P_{{\rm \leftarrow}})$.
This is the result quoted in Section~\ref{subsec:Few-atom-antenna}.

More precisely, for a two-layer antenna with phases chosen to emit
into a Gaussian mode with a beam waist $w_{0}$, 
$J_{j}\sim e^{ik_{0}z_{j}}LG_{0}^{0}(\vec{\rho}_{j},z_{j}),$
as discussed in Section~\ref{subsec:Few-atom-antenna}, one can similarly show
that the effective optical depth for forward emission reads 
$\mathcal{O}_{d}^{{\rm eff}}=8+32(w_{0}^{4}/\sigma^{2}),$
where $\sigma=3\lambda_{0}^{2}/(2\pi)$ is the scattering cross section
of a two-level atom and the calculation is performed for the interlayer
spacing $\delta_{z}=(3/4)\lambda_{0}$.

\section{Derivation of Chiral Master Equation for Two Nodes}\label{sec:chiralmasterequationderiviation}

Here we derive the chiral master equation for two interacting distant quantum nodes of master atom and antenna. We adiabatically eliminate the field and the antenna atoms as an effective reservoir for the master atoms, and obtain a `chiral' master equation, assuming a vacuum initial state for the photonic field. In Sec.~\ref{sec:derivationeffectiveME} we first sketch
the formalism, while in Sec.~\ref{sec:emergentchiralME} we provide explicit expressions.
\subsection{Derivation of effective master equation}\label{sec:derivationeffectiveME}

In a frame rotating with the optical frequency $\omega_0$, and eliminating the photonic field dynamics in a Born-Markov approximation, the model can be described by a master equation for the atomic density matrix $\tilde\rho$ describing the master atoms and the antenna atoms, reading
\begin{equation}
\begin{aligned}
\label{eq:ME2nodes}
\frac{d\tilde\rho}{dt}= -i\Big[H_\mathrm{int}  +\sum_{j,k=1}^{2N_{\rm a}}&\mathrm{Re}({H}_{nh})_{j,k}\sigma_j^+\sigma^-_k,\tilde\rho\Big]  \\
-2\sum_{j,k=1}^{2N_{\rm a}}&\mathrm{Im}({H}_{nh})_{j,k}\mathcal{D}[\sigma_{j}^{+},\sigma_{k}^{-}]\tilde\rho.
\end{aligned}
\end{equation}
Here, we integrated the dynamics of the photonic field, which we assumed initially in the vacuum state (zero temperature), we defined $H_\mathrm{int}\equiv\sum_{a,j}\left(J_{j,a}\sigma_{j}^{+}s_{a}^{-}+\text{h.c.}\right)$ and \mbox{$\mathcal{D}[\sigma_{j}^{+},\sigma_{k}^{-}]\tilde\rho\equiv\sigma_{k}^{-}\tilde\rho\sigma_{j}^{+}-\frac{1}{2}\{\tilde\rho,\sigma_{j}^{+}\sigma_{k}^{-}\}$}, and we recall that $H_{nh}\equiv-\Delta\mathbb{I}-i(\gamma_{e}/2)\left(\mathbb{I}+\mathbb{G}\right)$.
Following the notations of Ref.~\cite{Reiter2012}, we express Eq.~\eqref{eq:ME2nodes} in the
form
\begin{align}
\frac{d\tilde\rho}{dt}  =&-i\left(H_{\text{NH}}+V_{+}+V_{-}\right)\tilde\rho+i\tilde\rho\left(H_{\text{NH}}^{\dagger}+V_{+}+V_{-}\right)\nonumber \\
 & +\sum_{j,k=1}^{2 N_{\rm a}}\Gamma_{j,k}\left(L_{k}\tilde\rho L_{j}^{\dagger}\right),\label{eq:ME_1}
\end{align}
with the jump operators $L_{i}\equiv\sigma^-_{i}$, the collective emission rates given by the matrix $\Gamma_{j,k} \equiv -2\text{Im}(H_{nh})_{j,k}$, the non-hermitian Hamiltonian describing the dynamics of excitations in the arrays $H_{\text{NH}} \equiv\sum_{i,j}({H_{nh}})_{i,j}\sigma_{i}^{+}\sigma_{j}^{-}$, the term coupling excitations in the qubits to excitations in the arrays
\begin{align}
V_{+} & \equiv{\cal Q}\sum_{a=1}^{2}s_{a}^{-}\sum_{j=1}^{2N_{\rm a}}J_{j,a}\sigma_{l}^{+}{\cal P},\label{eq:V_plus}
\end{align}
and $V_{-} \equiv V_{+}^{\dagger}$.
Here $\mathcal P$ and $\mathcal Q$ are projectors defined as \mbox{${\cal P}\equiv\otimes_{a=1}^{2} \mathbb{I}_{a}\otimes_{j=1}^{2N_{\rm a}}\left\vert g\right\rangle _{j}\left\langle g\right\vert $} and
${\cal Q}\equiv\mathbb{I}-{\cal P}$, where we assume that at all time at most a single excitation is present in the atomic arrays. From the fact that $\Gamma_{j,k}$ is a real and positive-definite matrix, it can be diagonalized with an orthogonal
matrix $U$ as $\sum_{j,k}U_{l,j}^{T}\Gamma_{j,k}U_{k,l^{\prime}}=\Gamma_{l}\delta_{l,l^{\prime}}$, with $\Gamma_l\geq0$. 
The last term of Eq.~\eqref{eq:ME_1} then becomes

\begin{align*}
\sum_{j,k=1}^{2N_{\rm a}}\Gamma_{j,k}\left(L_{k}\tilde\rho L_{j}^{\dagger}\right) & =\sum_{j,k,l=1}^{2N_{\rm a}}U_{j,l}\Gamma_{l}U_{l,k}^{T}\left(L_{k}\tilde\rho L_{j}^{\dagger}\right)\\
 & =\sum_{l=1}^{2N_{\rm a}}\Gamma_{l}\left(\tilde{L}_{l}\tilde\rho\tilde{L}_{l}^{\dagger}\right),
\end{align*}
where we defined new jump operators as \mbox{$\tilde{L}_{l}\equiv\sum_{j}U_{l,j}L_{j}$}.
We now eliminate adiabatically the degrees of freedom for the atomic arrays, assuming
the weak couplings \mbox{$J_{j,a}\ll\left|\Delta-i{\gamma_{e}}/{2}\right|$},
such that the population of the antenna atoms is small at all times
(i.e. $\text{Tr}\left[Q\tilde\rho\right]\ll1$). 

Applying second order perturbation theory, the projected
density matrix ${\rho}\equiv{\cal P}\tilde\rho{\cal P}$ obeys the
Lindblad master equation 
\begin{equation}
\frac{d{\rho}}{dt}=-i\left[H_{\text{eff}},{\rho}\right]+\sum_{l=1}^{2N_{\rm a}}\Gamma_{l}{\cal D}\left[L_{\text{eff}}^{\left(l\right)\dagger},L_{\text{eff}}^{\left(l\right)},\rho\right],\label{eq:ME_eff}
\end{equation}
with the effective Hamiltonian and jump operators defined as
\begin{align}
H_{\text{eff}} & =-\frac{1}{2}V_{-}\left(H_{\text{NH}}^{-1}+\left(H_{\text{NH}}^{-1}\right)^{\dagger}\right)V_{+},\label{eq:H_eff}\\
L_{\text{eff}}^{\left(l\right)} & =L_{l}H_{\text{NH}}^{-1}V_{+}.\label{eq:L_eff_def}
\end{align}
Below we provide explicit expressions for Eqs.~\eqref{eq:H_eff} and \eqref{eq:L_eff_def}.

\subsection{Emergent chiral master equation}\label{sec:emergentchiralME}

Substituting the expressions of Eq.~(\ref{eq:V_plus})
in Eq.~(\ref{eq:H_eff}) we first obtain

\begin{align}
H_{\text{eff}} & =-{\cal P}\sum_{a,a^{\prime}=1}^{2} \sum_{j,k=1}^{2N_{\rm a}}J_{j,a}^*\text{Re}\left[H_{nh}^{-1}\right]_{j,k}J_{k,a'} s_{a}^{+}s_{a^{\prime}}^{-}{\cal P}.\label{eq:H_eff-1}
\end{align}
Analogously, we have for last term of Eq.~\eqref{eq:ME_eff}
\begin{align}
 & \sum_{l=1}^{2N_{\rm a}}\Gamma_{l}\mathcal D\left[\left(L_{\text{eff}}^{\left(l\right)}\right)^{\dagger},L_{\text{eff}}^{\left(l\right)},{\rho}\right]\nonumber \\
 & =2\sum_{a,a^{\prime}=1}^{2}\sum_{j,k=1}^{2N_{\rm a}}J_{j,a}^*\text{Im}\left[H_{nh}^{-1}\right]_{j,k}J_{k,a'}\mathcal D\left[s_{a}^{+},s_{a^{\prime}}^{-},{\rho}\right],
\end{align}
where we used the relation 
\begin{equation}
\sum_{i,i'=1}^{2N_{\rm a}}\left({H_{nh}^{-1}}\right)^{*}_{i,j}\Gamma_{i,i^{\prime}}\left({H_{nh}^{-1}}\right)_{i',k}=2\text{Im}\left[H_{nh}^{-1}\right]_{j,k}.
\end{equation} 
Finally, the master equation for the qubit reduced density matrix $ \rho$ reads
\begin{align}
\frac{d{\rho}}{dt}=  \sum_{a,a^{\prime}=1}^{2}\sum_{j,k=1}^{2N_{\rm a}}\Big(&i J_{j,a}^*\left[{H_{nh}^{-1}}\right]_{j,k}J_{k,a'}s_{a}^{+}s_{a^{\prime}}^{-}{\rho} \nonumber\\
  -&iJ_{j,a}^*\left[{H_{nh}^{-1}}\right]_{j,k}^{*}J_{k,a'}{\rho}s_{a}^{+}s_{a^{\prime}}^{-} \label{eq:ME_eff-1} \\
  +&2{J^*_{j,a}}\text{Im}\left[{H_{nh}^{-1}}\right]_{j,k}J_{k,a'}s_{a^{\prime}}^{-}\rho s_{a}^{+}\Big).\nonumber
\end{align}
This master equation can finally be expressed in the form of the main text by identifying $\gamma_{1,2}, \gamma_{L,R}$ and $\phi_{L/R}$.

\section{Deterministic Quantum State Transfer Protocols}\label{sec:pulseshapingQST}

In this section we provide expressions for the functions $f_{1,2}(t)$ realizing
Quantum State Transfer. Their explicit form can be obtained by
requiring the temporal shape of photons emitted by the first array
to be symmetric under time reversal, such that $f_{2}(-t)=f_{1}(t)$
is a solution. This is discussed in more details for example in Refs.~\cite{Cirac1997,Stannigel2011}.
We will assume for simplicity a symmetric scenario, where $\gamma_{a}\equiv \gamma$, and $\gamma_{R/L,a}\equiv \gamma_{R/L}$. In our simulations we use $f_{1}(t)=\sqrt{e^{\gamma t}/(2-e^{\gamma t})}$
for $t<0$, and $f_{1}(t)=1$ for $t\geq0$, although the results do not depend on the explicit expression of these
shapes.

For an initial pure state with a single excitation, the qubit density matrix
can be written as 
$
\rho(t)=\ket{\psi(t)}\bra{\psi(t)} + P_g(t)\ket{GG}\bra{GG},$
where 
\mbox{$\ket{\psi(t)}=c_{1}(t)\ket{EG}+c_{2}(t)\ket{GE}$}
with $c_1(t)$ and $c_2(t)$ the excitation amplitudes of the first and second qubit.
We get, assuming $\gamma_{L}\ll\gamma_{R}$, 
\begin{equation}
\begin{aligned}
\frac{dc_{1}}{dt}= & -  \frac{1}{2}\gamma f_{1}(t)^{2}c_{1}(t)\\
\frac{dc_{2}}{dt}= & -  \frac{1}{2}\gamma f_{2}(t)^{2}c_{2}(t)-\gamma_{R}e^{i\phi_R}f_{1}(t)f_{2}(t)c_{1}(t).
\end{aligned}
\end{equation}
Denoting here $2T$ the duration of the protocol from the initial time
$-T$ to the final time $T$, these equations can be integrated to
yield 
\mbox{$\left|c_{2}(T)\right|^{2}=\left({\gamma_{R}}/{\gamma}\right)^{2}\left[1-\mathcal{O}(e^{-\gamma T})\right]\left|c_{1}(-T)\right|^{2}$}.
The fidelity for QST is the success probability of transfer
for the initial condition $c_{1}(-T)=1$ and \mbox{$c_2(-T)=P_g(-T)=0$}, and is thus identified as $(\gamma_{R}/\gamma)^{2}$,
provided $\gamma T$ is taken large enough. In the simulations we
use $\gamma T=20$.

\section{Derivation of Quantum Stochastic Schr\"odinger Equation}\label{app:qsse}

In this Section we provide details omitted in Sec.~\ref{sec:QSSE},
and provide a QSSE formulation of the dynamics in the limit where
the antenna can be adiabatically eliminated. We consider only the case of two quantum nodes (\mbox{$a=1,2$}) of master atom and antenna, and the case of a single node is obtained by dropping the subscript $a$ everywhere.

\subsection{Adiabatic elimination of antenna atoms}

We consider here a minimal network of two master atoms ($a=1,2$),
with ground states $\left\vert G\right\rangle _{a}$ and excited states
$\left\vert E\right\rangle _{a}$, and with $s_{a}^{-}\equiv\left\vert G\right\rangle _{a}\!\left\langle E\right\vert $.
Each master atom is now coupled to a quantum
antenna consisting of $N_{a}$ atoms located at positions $\vec{r}_{i}$,
with ground states $\left\vert g\right\rangle {}_{i}$ and excited
states $\left\vert e\right\rangle _{i}$, with $\sigma_{i}^{-}\equiv\left\vert g\right\rangle _{i}\!\left\langle e\right\vert $
(where $i=1,2,\dots\cdots,2N_{a}$). In an interaction picture, the interaction Hamiltonian then reads
\begin{align}
V_{\mathrm{full}}(t)= & \sum_{i,a}\left(e^{-i\Delta t}J_{i,a}\sigma_{i}^{+}s_{a}^{-}+\text{h.c.}\right)\\
 & -d\sum_{i}\left(e^{-i\Delta t}\sigma_{i}^{+}\vec{p}^{*}\vec{\mathcal{E}}^{(+)}(\vec{r}_{i},t)+\text{h.c.}\right).\nonumber 
\end{align}
Assuming
the detuning defines the fastest timescale in the system (i.e.,~$\Delta\gg\gamma_{e},\delta\omega$ with
 $\delta\omega$ the bandwidth of the photonic field), we can eliminate the antenna atoms adiabatically, assuming they remain in their ground state at all time. The
evolution of the resulting system, effectively coupling the master atoms to
the photonic field, is governed by the following effective Hamiltonian
\begin{align}
V(t)= & \sum_{i,a}\frac{\left|J_{i,a}\right|^{2}}{\Delta}s_{a}^{+}s_{a}^{-}\nonumber\\
 & +\frac{d^{2}}{\Delta}\sum_{i}\left( \vec{p}\vec{\mathcal{E}}^{(-)}(\vec{r}_{i},t)\right) \left( \vec{p}^{*}\vec{\mathcal{E}}^{(+)}(\vec{r}_{i},t)\right) \nonumber\\
 & -\frac{d}{\Delta}\sum_{i,a}J_{i,a}^{*}s_{a}^{+}\left( \vec{p}^{*}\vec{\mathcal{E}}^{(+)}(\vec{r}_{i},t)\right) -\text{h.c.} \label{eq:Int_pic}
\end{align}
The first line represents a Stark shift redefining the master atom transition frequency, which can be compensated e.g. with additional AC Stark shifts. The second line represents an effective refraction index, which can be neglected in the limit $\gamma_e/\Delta \ll 1$. Finally, the third line contains the interaction term we are interested in.

\subsection{QSSE for the full 3D field}
We first rewrite the Hamitonian of Eq.~\eqref{eq:Int_pic} using the expansion of Eq.~\eqref{eq:expE} as
\begin{equation}\label{eq:Vtomegaqsse}
V(t)=i\int d\omega\left( e^{i(\omega-\omega_{0})t}\sum_{a}\kappa_a(\omega)b_{a,\omega}^{\dagger}s_a^{-} - \text{h.c.}\right).
\end{equation}
Here we replaced $\int d^3 k\to\int d\omega \, \omega^2 \int d\Omega/c^3$, with $d\Omega$ the differential solid angle, and we defined the coupling $\kappa_a(\omega)$
\begin{align}
\kappa_a(\omega)=&\sqrt{\frac{d^{2}\omega^{3}}{2c^{3}(2\pi)^{3}\Delta^2\epsilon_{0}}}
\\ \times&\sqrt{\sum_{i,j}J_{i,a}J_{j,a}^* \frac{8\pi}3\left(\delta_{i,j}+\text{Re}\left(\vec p^*\hat G(\vec r_{i}-\vec r_{j})\vec p\right)\right)},\nonumber
\end{align}
where we used 
\begin{equation}
\text{Re}\left(\vec p^*\hat G( r)\vec p\right)=\frac{3}{8\pi}\sum_{\lambda}\int d\Omega|\vec{p}\cdot\hat{e}_{\lambda,\vec k}|^{2}e^{-i\vec{k}\cdot\vec r}.
\end{equation}
The photon annihilation operators on the other hand are defined as
\begin{equation}\label{eq:baomega}
b_{a,\omega}=\frac{\omega\sum_{\lambda}\int d\Omega b_{\vec k}  (\vec{p}\cdot\hat{e}_{\lambda,\vec k})\sum_{i}e^{i\vec{k}\cdot\vec{r}_{i}}{J}^*_{i,a}}
{\sqrt{c^{3}\sum_{\lambda}\int d\Omega|\vec{p}\cdot\hat{e}_{\lambda,\vec k}|^{2}|\sum_{i}e^{i\vec{k}\cdot\vec{r}_{i}}{J}^*_{i,a}|^{2}}},
\end{equation}
such that they satisfy bosonic commutation relations \mbox{$[b_{a,\omega},b^\dagger_{a,\omega'}]=\delta(\omega-\omega')$}.

In order to obtain a QSSE from Eq.~\eqref{eq:Int_pic}, we need to perform a Born-Markov approximation and assume $\kappa(\omega)\approx\kappa(\omega_0)$. This requires in particular that \mbox{$\omega L_\perp/c\ll1$}, where $L_\perp$ denotes the spatial extent of the antenna, such that the phase factor $e^{ikL_\perp}$ acquired by a photon propagating in the antenna can be approximated by $e^{ik_0L_\perp}$. Finally, we define the quantum noise operators 
\begin{equation}\label{eq:batdef}
b_a(t)=\frac{1}{\sqrt{2\pi}}\int d\omega e^{-i(\omega-\omega_{0})t}b_{a,\omega},
\end{equation}
satisfying $[b_a(t),b^\dagger_{a}(t')]=\delta(t-t')$, and the Hamiltonian now reads 
\begin{equation}\label{eq:Vtsimple}
V(t)=i\sum_a\left(g_a b_a^{\dagger}(t)s_a^{-}-\text{h.c.}\right),
\end{equation}
with $g_a\equiv \sqrt{2\pi} \kappa_a(\omega_0)$ which is equivalent to the expression of Eq.~\eqref{eq:gamma_tot_large_Delta} if we identify $\gamma_{\text{tot},a}=|g_a|^2$.

\subsection{QSSE for the paraxial part of the field}
In order to obtain an effective 1D description of the dynamics, we decompose the QSSE obtained above into a paraxial part, with both antennas coupling to the same modes, and a non-paraxial part, where each antenna couple to its own bath (corresponding to the imperfections from $\beta<1$). Considering first only right-propagating modes, we project the electric field operator on a paraxial basis, and obtain, with $\vec r\equiv(\vec\rho,z)$, 
\begin{equation}
E^R_\text{par}(\vec \rho,z,t)=i\int dk \sum_n u_n(k,\vec \rho,z)e^{ikz}E^R_{n}(k)e^{-i(\omega_{k}-\omega_{0})t},
\end{equation} 
where $u_n(k,\vec \rho,z)$ forms an orthonormal basis for paraxial modes with momentum $k$ propagating along $z$ in the right direction, and 
\begin{equation}
E^R_{n}(k)=2\pi\int d^{2}q\sqrt{\frac{\omega_{k}}{2(2\pi)^{3}\epsilon_{0}}}\,b_{\vec{k}}\,v_{n}^{*}(k,\vec{q},z=0),
\end{equation}
where we defined the Fourier transform as \mbox{$
u_{n}(k,\vec{r},z)\equiv({1}/{2\pi})\int d^{2}qe^{i\vec{q}\cdot\vec{r}}v_{n}(k,\vec{q},z).$} Here we made the assumption that the transverse spectrum can be restricted to $|\vec q|\ll k_0$ (paraxial approximation).

The Hamiltonian for the interaction between master atom and paraxial field can then be written as
\begin{align}\label{eq:Vpart}
&V^R_\text{par}(t)\equiv \frac{d}{\Delta}\sum_{i,a}\left({E^R_\text{par}(\vec \rho_{i},z_{i},t)}^\dagger J_{i,a}s_a^-+\text{h.c.}\right)
\\ &= i \int d\omega \left(e^{i(\omega-\omega_{0})t}\sum_{a,n}\kappa^R_{n,a}(\omega)e^{-i\omega z_a/c}{b^R_{n,\omega}}^{\dagger}s_a^{-}-\text{h.c.}\right),\nonumber
\end{align}
where $z_a$ denotes the geometric center position of antenna $a$ along $z$, the coupling expresses as
\begin{equation}\label{eq:kappana}
\kappa^R_{n,a}(\omega)=\sqrt{\frac{d^{2}\omega}{4\pi c\epsilon_{0}\Delta^{2}}}\sum_{a,i}e^{-i\omega (z_{i}-z_a)/c}u_{n}^{*}(\omega,\vec{r}_{i},z_{i}){J}_{i,a}
\end{equation}
and the photon annihilation operator as
\begin{equation}\label{eq:bnomega}
b^R_{n,\omega}=\frac{1}{\sqrt{c}}\int d^{2}q\,b_{{\omega/c},\vec{q}}\,u_{n}^*(\omega,\vec{q},z=0)
\end{equation}
satisfying $[b^R_{n,\omega},{b^R_{n',\omega'}}^\dagger]=\delta(\omega-\omega')\delta_{n,n'}$.
We now perform a Born-Markov approximation, where we assume that $\kappa^R_{n,a}(\omega)\approx\kappa^R_{n,a}(\omega_0)$. Defining the quantum noise operators 
\begin{equation}\label{eq:bntdef}
b^R_{n}(t)=\frac{1}{\sqrt{2\pi}}\int d\omega e^{-i(\omega-\omega_{0})t}b^R_{n,\omega},
\end{equation}
which satisfy $[b^R_n(t),{b^R_{n'}}^\dagger(t')]=\delta_{n,n'}\delta(t-t')$, the Hamiltonian of Eq.~\eqref{eq:Vpart} now expresses as 
\begin{equation}
V^R_\text{par}(t)=i \sum_{a,n}g^R_{n,a}\left(e^{-i\omega_0 z_a/c}\,{b^R_{n}}^{\dagger}(t-z_a/c)s_a^{-}-\text{h.c.}\right),
\end{equation}
with $g^R_{n,a}\equiv \sqrt{2\pi} \kappa^R_{n,a}(\omega_0)$, which is equivalent to the expression of Eq.~\eqref{eq:gamma_n_large_Delta} if we identify the decay rate as $\gamma^R_{n,a}=|g^R_{n,a}|^2$.

For left-propagating paraxial modes, we perform a similar procedure, and we define all corresponding variables by replacing the superscript $R\to L$. We use a similar decomposition for the paraxial modes, with the waist located at the same position as for right-propagating modes, which is obtained by replacing the mode expressions as \mbox{$u_{n}\left(k,\vec{\rho},z\right)e^{ikz} \to u_{n}^{*}\left(k,\vec{\rho},z\right)e^{-ikz}$}, and finally obtain
\begin{equation}
V^L_\text{par}(t)=i \sum_{a,n}g^L_{n,a}\left(e^{i\omega_0 z_a/c}\,{b^L_{n}}^{\dagger}(t+z_a/c)s_a^{-}-\text{h.c.}\right).
\end{equation}

\subsection{Field decomposition}

From the mode definitions in Eqs.~\eqref{eq:batdef} and \eqref{eq:bntdef}, we can now decompose the modes interacting with the antennas into their paraxial and non-paraxial parts. 
Expanding the photon operators in terms of Eqs.~\eqref{eq:baomega} and \eqref{eq:bnomega}, we obtain
\begin{align}
[b_a(t),e^{-i\omega_0z_a/c}{b^R_n}^\dagger(t'-z_a/c)]&=\frac{\left(g^R_{n,a}\right)^*}{g_a}\delta(t-t'),
\\ 
[b_a(t),e^{i\omega_0z_a/c}{b^L_n}^\dagger(t'+z_a/c)]&=\frac{\left(g^L_{n,a}\right)^*}{g_a}\delta(t-t'),
\end{align}
where we identify $|g^R_{n,a}/g_{a}|^2$ as the Purcell $\beta$-factor. Finally, we define the coupling to unwanted non-paraxial modes as
$g_a'\equiv \sqrt{|g_{a}|^2 - \sum_n \left(|g^R_{n,a}|^2+|g^L_{n,a}|^2\right)}.$
This provides a definition for the annihilation operator of photons in these unwanted modes $b_a'(t)$ as
\begin{align}
b_a(t)\equiv &\sum_n\frac{\left(g^R_{n,a}\right)^*}{g_a}e^{i\omega_0z_a/c}b^R_n(t-z_a/c)
\\ +& \sum_n\frac{\left(g^L_{n,a}\right)^*}{g_a}e^{-i\omega_0z_a/c}b^L_n(t+z_a/c)+\frac{g_a'}{g_a} b_a'(t),\nonumber
\end{align}
which we assume independent with \mbox{$[b_1(t),{b_2'}^\dagger(t')]=[b^{R/L}_n(t),{b_a'}^\dagger(t')]=0$}, such that $[b_a'(t),{b_{a'}'}^\dagger(t')]=\delta(t-t')\delta_{a,a'}$, and from Eq.~\eqref{eq:Vtsimple} we obtain Eq.~\eqref{eq:QSSE_V2def}, where we set $z_1=0$ and $z_2=2z_0$.

\section{Details on the Rydberg Implementation\label{sec:Details-of-Rydberg}}

In this Section we provide details on our Rydberg implementation discussed
in Sec.~\ref{sec:Atomic-Implementations}.

\subsection{Model}

In a frame rotating with the laser frequencies, the quantum-optical
Hamiltonian describing our model can be written in the form $H_{\mathrm{Ryd}}=H_{0A}+H_{AF}+H_{0F}+H_{\mathrm{losses}}$,
where we have

\begin{eqnarray*}
H_{0A} & = & -\sum_{i}\left[(\Delta_{d}+\tilde{\Delta})\ket{R}_{i}\bra{R}\right.\ \\
 &  & \quad\quad+\left.\Delta_{d}\ket{R'}_{i}\bra{R'}+\Delta_{c}\ket{e}_{i}\bra{e}\vphantom{{\tilde{\Delta}}}\right]\\
 & + & \sum_{i}\left[\Omega_{d}(\vec{r}_{i})\ket{g}_{i}\bra{R'}+\Omega_{c}(\vec{r}_{i})\ket{R}_{i}\bra{e}+\mathrm{h.c.}\right]\\
 & + & \sum_{i}V_{\mathrm{dd}}(\vec{r_{i}}-\vec{r}_{m})\left(\ket{ER'_{i}}\bra{GR_{i}}+\mathrm{h.c}\right)\\
 & + & \sum_{i<j}V'_{\mathrm{dd}}(\vec{r}_{i}-\vec{r}_{j})\left(\ket{R_{j}R'_{i}}\bra{R'_{j}R_{i}}+\mathrm{h.c}\right),
\end{eqnarray*}
with $V_{\mathrm{dd}}(\vec{r}_{i}-\vec{r}_{m})=C_{3}(1-3\cos^{2}\theta_{i,m})/r_{i,m}^{3}$
the desired `flip-flop' process transferring excitations between master
atom and antenna atoms, and where $V'_{\mathrm{dd}}(\vec{r_{i}}-\vec{r}_{j})=C'_{3}(1-3\cos^{2}\theta_{ij})/r_{i,j}^{3}$
describes dipole-dipole couplings between antenna atoms. Here, $\vec{{r}}_{m}$
is the position of the master atom, $r_{ab}=|\vec{r}_{a}-\vec{r}_{b}|$,
and \mbox{$\cos\theta_{ab}=[(\vec{r}_{a}-\vec{r}_{b})\cdot\vec{z}\,]/r_{ab}$}.
For the levels chosen above, we have \mbox{$C_{3}\approx49.3\,h\textrm{\,MHz}\,\mu{\rm m}^{3}$},
$C_{3}'\approx41.5\,h\,\textrm{MHz}\,\mu{\rm m}^{3}$. The Hamiltonians
$H_{AF}$ and $H_{0F}$ are introduced in the main text. Finally,
we model in a first approximation the natural decay of the Rydberg
states with a non-Hermitian Hamiltonian \mbox{$H_{\mathrm{losses}}=-i(\gamma_{r}/2)\left(\ket{R}_{i}\bra{R}+\ket{E}_{i}\bra{E}\right)$},
with \mbox{$\gamma_{r}=2\pi\times3.6\,\text{KHz}$}.

\subsection{Perturbative regime}

By choosing $\Delta_{d}+\tilde{\Delta}=0$, and in the regime \mbox{$\Omega_{d},V_{\mathrm{dd}}\ll\Delta_{d}$},
we can eliminate in second-order perturbation theory the state $\ket{R'}_{i}$
and obtain 
\begin{eqnarray}
H_{0A} & = & \sum_{i}J_{i}\left(\ket{Eg_{i}}\bra{GR_{i}}+\mathrm{h.c}\right)\nonumber \\
 & + & \sum_{i<j}J'_{ij}\left(\ket{R_{j}g_{i}}\bra{g_{j}R_{i}}+\mathrm{h.c}\right)\nonumber \\
 & + & \sum_{i}\left[\Omega_{c}(\vec{r}_{i})\ket{R}_{i}\bra{e}+\mathrm{h.c.}\right]-\Delta_{c}\ket{e}_{i}\bra{e},\label{eq:H0a_threelevels}
\end{eqnarray}
allowing long-range coherent excitation transfer from master atom
to antenna atoms. Here the `dressed' couplings $J_{i}=V_{\mathrm{dd}}(\vec{r}_{i}-\vec{r}_{m})\Omega_{d}(\vec{r}_{i})/\Delta_{d}$
can be engineered via the dressing-laser Rabi frequency $\Omega_{d}(\vec{r}_{i})$.
Note that we did not write the additional AC Stark shifts contributions,
which can be included in the definitions of the detunings (or compensated
via additional laser couplings).

Note that our effective Hamiltonian \eqref{eq:H0a_threelevels} is
not identical to the model presented in the main text {[}see Eq.~\eqref{eq:H0A}{]}.
First, instead of two level atoms with detunings $\Delta$, we obtain
here an antenna built from three-level atoms, where the decay from
Rydberg to ground state is Raman assisted by the control laser $\Omega_{c}(\vec{r}_{i})$.
Second, excitations can also hop between antenna atoms with \mbox{$J'_{ij}=V'_{\mathrm{dd}}(\vec{r}_{i}-\vec{r}_{j})\Omega_{d}(\vec{r}_{i})\Omega_{d}^{*}(\vec{r}_{j})/\Delta_{d}^{2}$}.

\subsection{Study of three-level atoms antennas}

To assess the performance of three-level antennas, we calculate numerically
the spatial profile of the mode $\vec{\varphi}(\vec{r}\,)$ generated
via the excitation transfer to the antenna, as governed by Eq.~\eqref{eq:H0a_threelevels}.
We can also obtain an analytical expression for the decay rate $\gamma_{\mathrm{tot}}$
and for $\vec{\varphi}(\vec{r}\,)$. Assuming $\Delta_{c}=0$ and a
strong control field $\Omega_{c}\gg J_{i},J_{i,j}^{\prime},\gamma_{r}$,
we obtain Eqs.~\eqref{eq:gamma_n_large_Delta}, \eqref{eq:gamma_tot_large_Delta}
with the identification \mbox{$J_{i}/\Delta\rightarrow J_{i}/\Omega_{c}\left(\vec{r}_{i}\right)$}.
In this limit, we establish a direct connection between our Rydberg
implementation and our model of two-level antenna presented in the
main text.

\end{document}